\documentclass[10pt,reprint,amsmath,amssymb,amsfonts,twocolumn,showpacs,preprintnumbers,superscriptaddress,aps,pra]{revtex4-1}
\usepackage{graphicx}
\usepackage{multirow}
\usepackage{color}
\usepackage{colordvi}

\newtheorem{conj}{Conjecture}

\usepackage[bookmarks=false,pdfstartview={FitH}]{hyperref}

\begin{document}
\title{Multiple-spin coherence transfer in linear Ising spin chains and beyond:\\  numerically-optimized pulses and experiments}
\author{Manoj Nimbalkar}
\email{manoj.nimbalkar@tum.de}
\affiliation{Department Chemie, Technische Universit\"at M\"unchen, Lichtenbergstrasse~4, 85747 Garching, Germany}

\author{Robert Zeier}
\email{robert.zeier@ch.tum.de}
\affiliation{Department Chemie, Technische Universit\"at M\"unchen, Lichtenbergstrasse~4, 85747 Garching, Germany}

\author{Jorge L.~Neves}
\affiliation{Department Chemie, Technische Universit\"at M\"unchen, Lichtenbergstrasse~4, 85747 Garching, Germany}
\affiliation{Laboratorio de Genomica Estrutural, Instituto de Biof\'isica Carlos Chagas Filho, Universidade Federal do Rio de Janeiro, Rio de Janeiro RJ 21941-590, Brazil}

\author{S.~Begam Elavarasi}
\affiliation{Department Chemie, Technische Universit\"at M\"unchen, Lichtenbergstrasse~4, 85747 Garching, Germany}
\affiliation{B.~S.~Abdur Rahman University, Seethakathi Estate, Vandalur, Chennai 600048, India}

\author{Haidong Yuan}
\affiliation{Department of Mechanical Engineering, Massachusetts Institute of Technology, 
77 Massachusetts Avenue, Cambridge, MA 02139, USA}
\affiliation{School of Engineering and Applied Sciences, Harvard University, 
33 Oxford Street, Cambridge, MA 02138, USA}

\author{Navin Khaneja}
\affiliation{School of Engineering and Applied Sciences, Harvard University, 
33 Oxford Street, Cambridge, MA 02138, USA}

\author{Kavita Dorai}
\affiliation{IISER Mohali, MGSIPAP Complex, Sector~26, Chandigarh 160019, India}

\author{Steffen~J.~Glaser}
\email{steffen.glaser@tum.de}
\affiliation{Department Chemie, Technische Universit\"at M\"unchen, Lichtenbergstrasse~4, 85747 Garching, Germany}

\date{October 24, 2011}


\begin{abstract}
We study multiple-spin coherence transfers in linear Ising spin chains with nearest neighbor couplings. 
These constitute a model for efficient information transfers in future quantum computing devices and
for many multi-dimensional experiments for the assignment of complex spectra in nuclear magnetic resonance
spectroscopy. We complement prior analytic techniques 
for multiple-spin coherence transfers with a systematic numerical study where
we obtain strong evidence that a certain analytically-motivated family of restricted controls
is sufficient for time-optimality. In the case of a linear three-spin system, additional
evidence suggests that prior analytic pulse sequences using this family of restricted controls
are time-optimal even for arbitrary local controls. In addition, we compare the pulse sequences
for linear Ising spin chains to pulse sequences for more realistic spin systems with additional long-range couplings 
between non-adjacent spins. We experimentally implement the derived pulse sequences
in three and four spin systems and demonstrate that they are applicable in realistic settings
under relaxation and experimental imperfections---in particular---by deriving broadband pulse sequences
which are robust with respect to frequency offsets.
\end{abstract}

\maketitle

\section{Introduction\label{sec:intro}}

The  control of spin dynamics in chains of coupled spins-$1/2$ is a topic of both theoretical and practical interest \cite{LSM61, PUL96, GB91, LGB93, Glaser93, GQ96, White83, Mattis88, MBSHBE97, KG02, YGK07, YZK08}. 
On the one hand, the use of spin chains is considered for the efficient transfer of information in future quantum computing devices \cite{YY99, SP09}. On the other hand, coherence transfer between remote spins is the basis of many multi-dimensional experiments  for the assignment of complex spectra \cite{RGD89, EFGD90, CFPS96} in nuclear magnetic resonance  (NMR) spectroscopy.
In addition to linear spin chains with only nearest neighbor couplings, in realistic settings also long-range couplings between non-adjacent spins must be considered.
For example in  $^{13}\mathrm{C}$ and $^{15}\mathrm{N}$ labeled proteins, the nuclei in the protein backbone  form a chain of coupled spins-$1/2$ with dominant next neighbor  $^1J$ (single bond) couplings and smaller $^2J$  and $^3J$ couplings (via two or three chemical bonds)  between non-adjacent spins in the chain  \cite{CFPS96}.

Here we focus on the efficient creation of multi-spin operators from a single-spin operator in a spin chain, such as the creation of multiple-spin order from polarization of the first spin
\begin{equation}\label{transfer_z}
 I_{1z} \rightarrow   2^{n-1} I_{1z}  I_{2z}  \cdots I_{(n-1) z} I_{n z}.
\end{equation}
The transfer shown in Eq.~\eqref{transfer_z} is just a prototype example of a general transfer of the form
\begin{equation}\label{transfer_gen}
 I_{1\delta} \rightarrow   2^{n-1} I_{1\epsilon_1}   \cdots  I_{n \epsilon_n}, 
\end{equation}
where $\delta, \epsilon_k  \in \{x,  y,  z  \}$ for $k=1, \ldots ,n$.
Note that the transformations in Eqs.~\eqref{transfer_z} and \eqref{transfer_gen} are identical up to local spin rotations. Hence, in the limit where the time for selective rotations of individual spins is negligible 
(compared to $1/(2 J_{max})$, where $J_{max}$ is the largest spin-spin coupling  constant in the chain),
the transformations in Eqs.~\eqref{transfer_z} and \eqref{transfer_gen} can be achieved in the same amount of time. (This situation is typical for heteronuclear NMR experiments in liquid state, where the control amplitudes for single spin operators are orders of magnitude larger than the largest coupling constants.)
In Eq.~\eqref{transfer_gen}, the initial single-spin state is not limited to  longitudinal magnetization (polarization $I_{1z}$) but may also be 
 transverse magnetization (in-phase coherence  $I_{1x}$ or $I_{1y}$) \cite{CFPS96, EBW:1997}. Examples of multi-spin target operators in Eq.~\eqref{transfer_gen}  containing one or several transverse operators include states of the form
$2^{n-1} I_{1z}  I_{2z}  \cdots I_{(n-1) z} I_{n x}$ (corresponding to anti-phase coherence of spin $n$ with respect to spins $1$ to $n-1$), 
and
$2^{n-1} I_{1x}  I_{2x}  \cdots I_{(n-1) x} I_{n x}$ (corresponding to multi-quantum coherence), which are relevant in  so-called ``out and back" transfer schemes  \cite{CFPS96, AITB91, SSG99} and in the creation of multiple-quantum coherence  \cite{EBW:1997, LE85}, respectively.

We consider the case of Ising-type spin chains \cite{Isi:1925, Cas:1989}, corresponding to the weak coupling limit  \cite{EBW:1997}. In this limit, which is an excellent approximation for hetero-nuclear NMR experiments, the  coupling Hamiltonian for a pair of spins $k$ and $l$ has the form
$${H}^{\text{weak}}_{kl}= 2 \pi J_{kl} I_{kz}  I_{lz},$$
where $J_{kl}$ is  the coupling constant in units of Hertz (Hz).
In conventional experiments, the standard methods to achieve transfer in Eq.~\eqref{transfer_gen}  are based on COSY- or RELAY-type transfer steps \cite{EBW:1997, CFPS96}, which are realized in hetero-nuclear experiments by a series of INEPT building blocks \cite{MF79} (see Sec.~\ref{model}).
The transfer time is determined by the size of the coupling constants $J_{kl}$ in a given spin system. For example, in a linear spin chain with only next-neighbor couplings, the total duration is given by 
$$T_{\text{conv}}= (J^{-1}_{12}  + J^{-1}_{23}  + \cdots   + J^{-1}_{(n-1)n}  )/2.$$
We are interested in finding the shortest possible time to achieve the transfer in Eq.~\eqref{transfer_gen} or conversely, the maximum transfer amplitude for any
given time, which remains an open question up to now.

For relatively simple spin systems, consisting of up to three spins,  time-optimal  \cite{KBG:2001, RKG:2002, KGB:2002, KKG:2005, KHSY:2007, FYSG09, ALZBGS10} and relaxation-optimized \cite{KRLG03, KLG03, SKG04, FILWGK05, SKG05, LZBGS10} pulse sequences have been recently found analytically based on methods of optimal control theory \cite{NKGK10,SRLKG03, KLKG05, KSKGB08, GSBKNLG08}, establishing rigorous physical limits of minimal transfer times or minimal relaxation losses, respectively. In addition to powerful analytical tools, optimal control theory also provides efficient numerical algorithms for the optimization of pulse sequences, such as the gradient ascent pulse engineering (GRAPE) algorithm, exploiting the known equation of motion for the spin system \cite{KRKSHG05, TVKKGN09, MSGFGSS11, FSGK11}.  With this algorithm it is possible to optimize tens of thousands of pulse sequence parameters and the resulting pulse sequences are not limited to previously known transfer schemes. However, in contrast to analytical methods proving global optimality of a given pulse sequence, there is no guarantee that numerical optimal control algorithms like GRAPE will converge to the global optimum
\cite{PT11}. Never-the-less, in 
cases where the theoretical limits are known, the GRAPE algorithm closely approached these limits \cite{KRKSHG05, LZBGS10b}. This motivated  its use also in cases for which analytical results on the global optimum are presently unknown in order to explore the physical limits of the maximum possible transfer efficiency as a function of transfer time, resulting in so-called TOP (time-optimal pulse) curves \cite{KSKGL04, KSKGB08, NHRSKG06, pomplun:2008, PG10, BG10}.
Furthermore, additional effects such as relaxation \cite{GKLGS07, LZGS11}, radiation damping \cite{ZLSBG11}, and experimental constraints and imperfections---such as limited control amplitudes and control field inhomogeneities  \cite{SRLKG04, SKLBBKG06, SMWGG11}---can be taken into account to find highly robust pulses suitable for practical applications under realistic conditions.

Assuming a restricted pulse structure (see Secs.~\ref{sec:lin3spin_theory} 
and \ref{sec:3_4_spin_theory}) analytic pulses were derived 
in Refs.~\cite{YGK07} and \cite{YZK08}, respectively,
for the case of equal and unequal couplings. This results in significantly shorter
transfer times compared to conventional approaches, however it was not clear
how closely the performance of the derived pulse sequences converges to the time-optimal performance.

In this work, we summarize the analytic approach 
of Refs.~\cite{YGK07,YZK08} (see Sections~\ref{sec:lin3spin_theory} and \ref{sec:3_4_spin_theory})
and explore its time-optimality by conducting a systematic numerical study 
of the considered coherence transfer
(see Sections~\ref{sec:lin3spin_numeric} and \ref{sec:3_4_spin_numeric}). Focusing on the
case of linear Ising spin chains with three and four qubits we compare
the duration of pulse sequences for arbitrary pulse structures with the
restricted pulse structure motivated by the analytic pulses.
We also numerically analyze linear Ising spin chains for
up to six spins (see Section~\ref{sec:longer_linspin}). In addition, our numerical approach  makes
it also possible to investigate more realistic spins systems with more general 
coupling topologies (see Section~\ref{sec:3_4_spin_loopedspin}).

In Sec.~\ref{sec:experiment}, we show how to make the pulse sequences robust with respect to off-resonance effects using the DANTE (Delays Alternating with Nutations for Tailored
Excitation) approach \cite{MF78, KLKLG04,  KHSY:2007}. Finally, we present experimental results for model spin chains consisting of three and four hetero-nuclear spins-$1/2$, demonstrating good performance of the new sequences under experimental conditions and comparing the results to conventional pulse sequences.

\section{Coherence transfer in linear Ising spin chains\label{model}}

Throughout this work we mostly consider linear Ising spin chains which
have only direct couplings between neighboring spins \cite{Isi:1925,Cas:1989}. (Later we 
will also allow additional couplings between non-neighboring spins.) Assume that a
chain of $n$ spins is placed in a static external magnetic field along the $z$-direction
and that neighboring spins are coupled by an Ising interaction where the coupling 
strengths $J_{\ell,\ell+1}$ are fixed but may depend on the position $1\leq \ell \leq 
n{-}1$ in the chain. Without any control, the system evolves freely under its drift Hamiltonian
$$
H_d=2~\pi~\sum_{l=1}^{n-1} J_{l, l+1} I_{lz}I_{(l+1)z}.
$$
The drift Hamiltonian is given in a suitably 
chosen multiple rotating frame, which rotates simultaneously at the resonance 
frequency of each spin. We use the product-operator basis 
$I_{\ell \nu}{=} \bigotimes_{j} I_{a_{j}}$ where $a_{j}{=}\nu$ for $j{=}\ell$ 
and $a_{j}{=}0$ otherwise  (see Ref.~\cite{EBW:1997}).
The matrices 
$I_{x}{:=}
\left(\begin{smallmatrix}
0 & 1 \\
1 & 0
\end{smallmatrix}
\right)/2$,
$
I_{y}{:=}
\left(
\begin{smallmatrix}
0 & -i \\
i & 0
\end{smallmatrix}
\right)/2
$, and
$
I_{z}{:=}
\left(
\begin{smallmatrix}
1 & 0 \\
0 & -1
\end{smallmatrix}
\right)/2
$ are the Pauli spin matrices
and
$
I_{0}{:=}
\left(
\begin{smallmatrix}
1 & 0 \\
0 & 1
\end{smallmatrix}
\right)
$
is the $(2\times 2)$-dimensional identity matrix.
In addition to the free evolution, we assume that individual spins can be 
selectively excited using radio-frequency (rf) pulses, which is the case if the 
Larmor frequencies of the spins are well separated as compared to the 
coupling strengths $J_{\ell,\ell+1}$. Thus controls on individual spins can be 
applied on a much faster time scale as compared to the free evolution 
w.r.t.\ the drift Hamiltonian.

We derive explicit controls for the amplitude and phase
of the external rf fields implementing a unitary evolution which transforms
an initial polarization $I_{1x}$ on the first spin to a multiple-spin 
state $2^{n-1}(\prod_{\ell=1}^{n-1}I_{\ell y})I_{nz}$ 
while minimizing the pulse duration $t_p$. In the following, we often 
compare control pulses with the \emph{conventional strategy}, which consists 
of $n{-}1$ steps of free evolution ($1\leq m \leq n{-}1$) $$2^{m-1}I_{1y}\cdots 
I_{m-1,y} I_{mx} \xrightarrow{H_d} 2^{m}I_{1y}\cdots I_{m y}I_{m+1,z}$$ where 
each individual step---besides the final one---is followed by one hard 
$\tfrac{\pi}{2}$-pulse on the $(m{+}1)$th spin along the $y$-direction. 
As each period of free evolution is of length $1/(2J_{\ell,\ell+1})$ 
where $J_{\ell,\ell+1}$ is given in Hz, the total evolution time 
is given by $t_p=\sum_{\ell=1}^{n-1} 1/(2 J_{\ell,\ell+1})$.

\section{Linear three-spin chains: analytic and numerical approaches\label{sec:lin3spin}}

\subsection{Analytic approach\label{sec:lin3spin_theory}}
\begin{figure}[tb]
	\includegraphics[width=0.5\columnwidth]{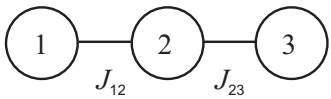}
	\caption{A linear three-spin chain has only direct couplings $J_{12}$ and $J_{23}$ between 
	neighboring spins.}
	\label{fig:ising3spinsytem}
\end{figure}

In this section,
we consider the model of Sect.~\ref{model} in the case of linear three-spin chains
(see Fig.~\ref{fig:ising3spinsytem}).
In the most general case, one could allow independent controls on each of the three 
spins along both the $x$- and $y$-direction. But in order to simplify the control problem
we allow only one control on the second spin along the $y$-direction.
This might not lead to time-optimal controls. But even
using this restricted model, controls which are shorter 
as compared to the conventional strategy
were obtained in Ref.~\cite{YZK08} (see also \cite{YGK07}).
In the following,
we summarize the analytical approach of Ref.~\cite{YZK08}.

\begin{figure}[tb]
	\includegraphics[width=0.49\columnwidth]{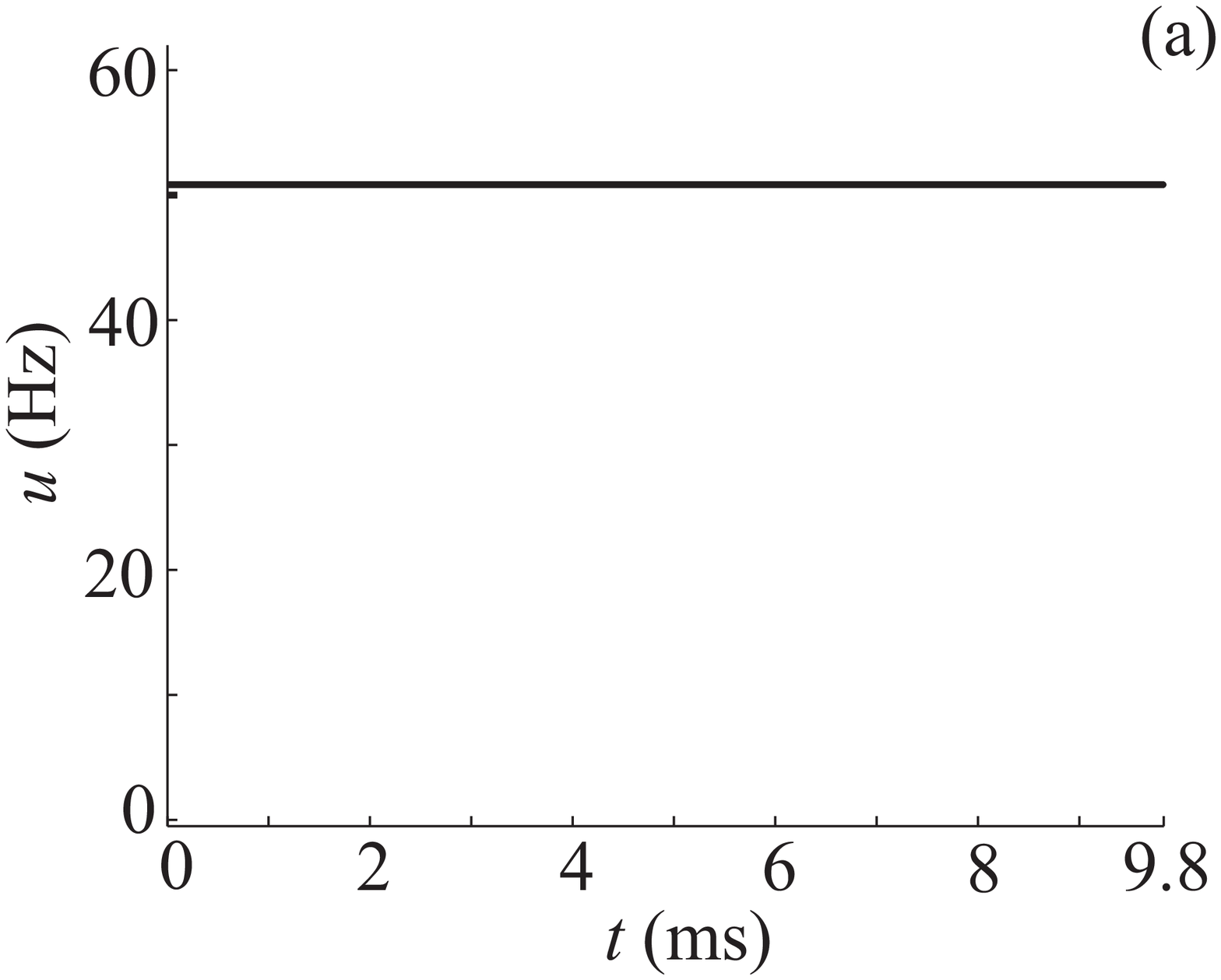} \includegraphics[width=0.49\columnwidth]{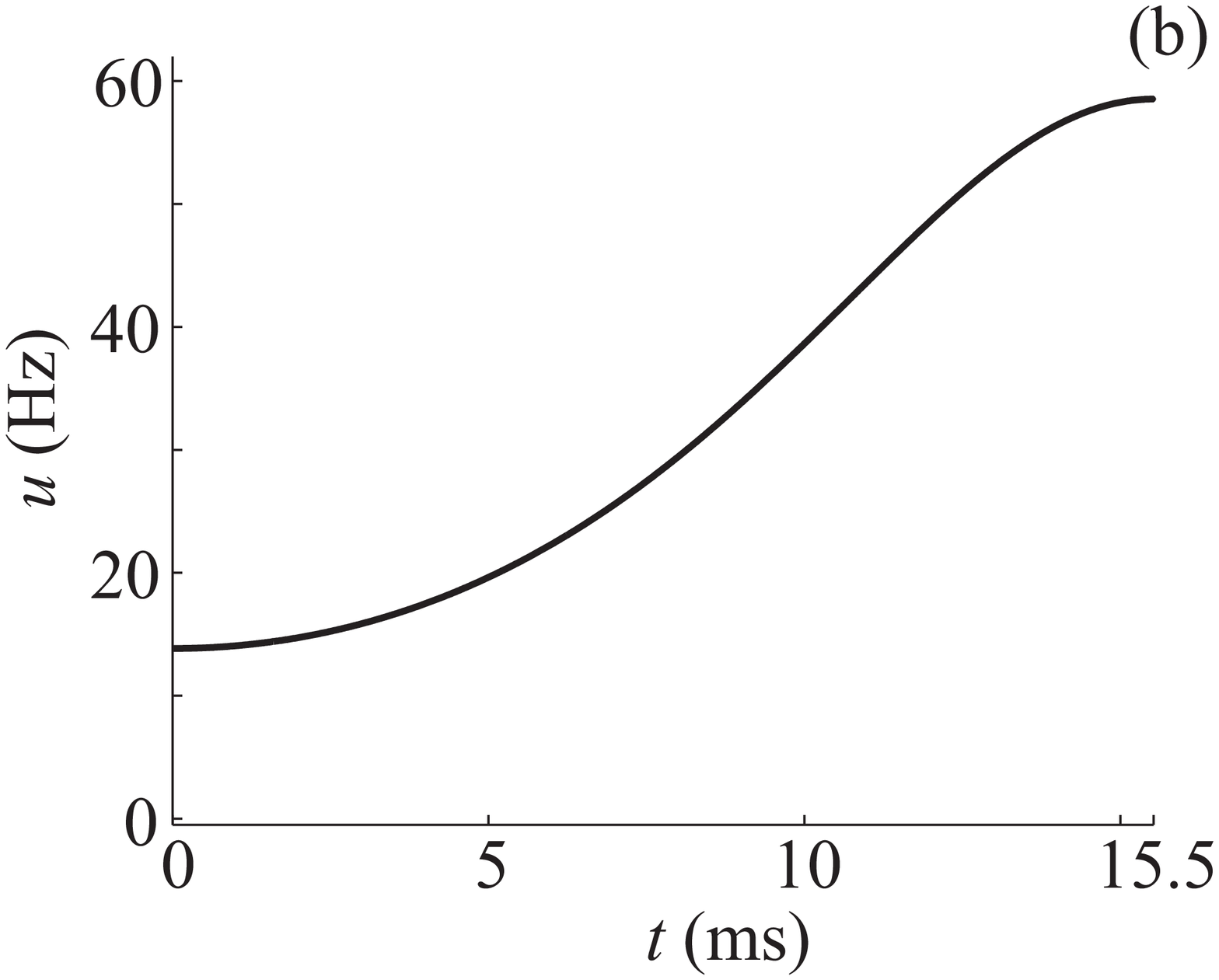}
	\caption{Analytic pulses for linear three-spin chains are given in the cases of (a) 
	$k=J_{23}/J_{12}=88.05/88.05=1$ and (b) $k=1.59 \approx
	J_{23}/J_{12}=73.1/46$.}
	\label{fig:ising3spin_algebraic_shape}
\end{figure}

Starting from an initial state $I_{1x}$ and using 
only one control on the second spin along the $y$-direction,
we can analyze the control problem on the subspace spanned
by the operators $I_{1x}$, $2I_{1y}I_{2z}$, $2I_{1y}I_{2x}$, and
$4I_{1y}I_{2y}I_{3z}$ as compared to the full $63$-dimensional space of operators.
Using the notation $\langle O\rangle:=\mathrm{Tr}(O \rho)$
for the expectation value and $\mathrm{Tr}$ for the trace,
we denote the corresponding expectation values
by $x_1=x_1(t)=\langle I_{1x}\rangle$, 
$x_2=x_2(t)=\langle 2I_{1y}I_{2z}\rangle$,
$x_3=x_3(t)=\langle 2I_{1y}I_{2x}\rangle$, and
$x_4=x_4(t)=\langle 4I_{1y}I_{2y}I_{3z}\rangle$. We obtain the differential equation
\begin{equation}\label{diff1}
\begin{pmatrix}
\dot{x}_1\\
\dot{x}_2\\
\dot{x}_3\\
\dot{x}_4
\end{pmatrix}
= \pi
\begin{pmatrix}
0 & -1 & 0 & 0 \\
1 & 0 & -u & 0 \\
0 & u & 0 & -k \\
0 & 0 & k & 0 
\end{pmatrix} 
\begin{pmatrix}
x_1\\
x_2\\
x_3\\
x_4
\end{pmatrix},
\end{equation}
where $u=u(t)$ denotes the amplitude of the control on the second spin along the $y$-direction
and $k=J_{23}/J_{12}$. Using the coordinates $(x_1,x_2,x_3,x_4)^T$
we aim to time-efficiently transfer $(1,0,0,0)^T$ to $(0,0,0,1)^T$. 

Now, we change from the coordinates $(x_1,x_2,x_3,x_4)^T$ to the coordinates 
$$(r_1,r_2,r_3)^T=(x_1,\sqrt{x_2^2+x_3^2},x_4)^T$$ on the sphere
where $\theta=\theta(t)$ is given by
$\tan\theta={x_3}/{x_2}$.
This transforms Eq.~\eqref{diff1} to
\begin{equation*}
\begin{pmatrix}
\dot{r}_1\\
\dot{r}_2\\
\dot{r}_3
\end{pmatrix}
= \pi
\begin{pmatrix}
0 & -\cos\theta & 0 \\
\cos\theta & 0 & -k\sin\theta \\
0 & k\sin\theta & 0
\end{pmatrix}
\begin{pmatrix}
r_1\\
r_2\\
r_3
\end{pmatrix}.
\end{equation*}
In the new coordinates, we want to time-efficiently transfer 
$(1,0,0)^T$ to $(0,0,1)^T$.

\begin{figure}[tb]
	\includegraphics[width=0.99\columnwidth]{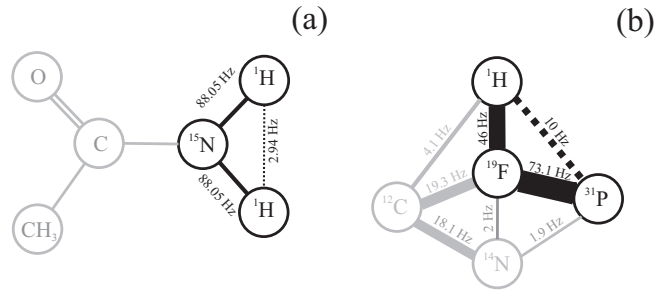}
	\caption{The schematic coupling topologies of (a) ethanamide and (b) diethyl-(dimethylcarbonyl)fluoromethylphosphonate (see \cite{pomplun:2010,MPBZEFG11}) 
	result in experimental
	three-spin systems with coupling ratios (a) $k=1=88.05/88.05$ and (b) $k=1.59\approx 73.1/46$.
	Larger couplings are shown as solid-black lines, and smaller couplings are shown as dashed-black lines. 
	Decoupled spins are given in gray color.}
	\label{fig:3_quibit}
\end{figure}

In order to find the time-optimal controls, Euler-Lagrange equations were set up
and solved in Ref.~\cite{YZK08} leading to the differential equation 
\begin{equation}\label{theta_eq}
\ddot{\theta} = \frac{k^2-1}{2} \sin2\theta
\end{equation}
for the variable $\theta$. The differential Eq.~\eqref{theta_eq} can be numerically integrated if
the initial values 
$\theta(0)$ and $\dot{\theta}(0)$ are known. Using the results of Ref~\cite{YZK08}
one can determine conditions on the initial values:
In the case of $(r_1(0),r_2(0),r_3(0))^T=
(1,0,0)^T$ one can deduce that $\theta(0)=0$, but $\dot{\theta}(0)$ is undetermined.
In Ref.~\cite{YZK08} combinations of one-dimensional searches were used to
determine the optimal $\theta_{\text{opt}}(t)$ and the
time-optimized control as $u_{\text{opt}}(t)=
J_{12} \dot{\theta}_{\text{opt}}(t)$. Examples for the corresponding
(semi-)analytic pulses are shown in
Fig.~\ref{fig:ising3spin_algebraic_shape}. The 
values are motivated by the experimental systems given in
Fig.~\ref{fig:3_quibit}.

\begin{table}[tb]
\caption{\label{tab:linear_3spin_k1_shape} 
We compare the duration $t_{p}$, the logarithmic fidelity $F$, and the shape of 
numerically-optimized pulses for a linear three-spin chain with coupling 
ratios (a) $k=1$ and (b) $k=1.59$. The number of controls $\#u$ is given 
in the first column. In the third 
column we present  the 
corresponding logarithmic  
TOP curves. The second column shows an example 
of a shaped pulse, whose position is denoted with an $\sf{x}$ in the 
logarithmic TOP curve. The rf control on the middle spin along the $y$-axis is 
plotted using a solid-black line. Other rf controls are plotted using 
dashed or solid-gray lines.}
\begin{ruledtabular}
		\begin{tabular}[t]{c  c  c}
		$\#{}u$ & pulse shape  & logarithmic TOP curve\\
		\hline \\[-3mm]
		\multicolumn{3}{l}{(a)~$k=1$:}\\
		 \raisebox{0.6cm}{1} &
		\includegraphics[width=0.45\columnwidth]{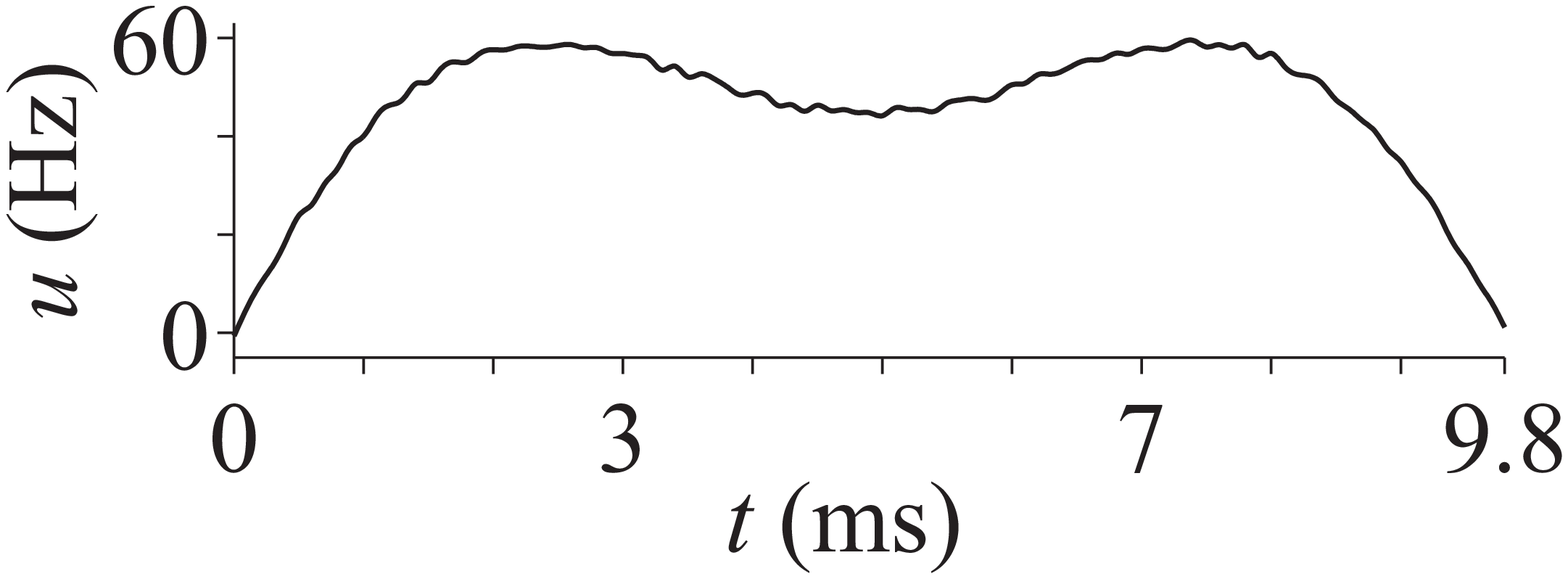} &
		\includegraphics[width=0.45\columnwidth]{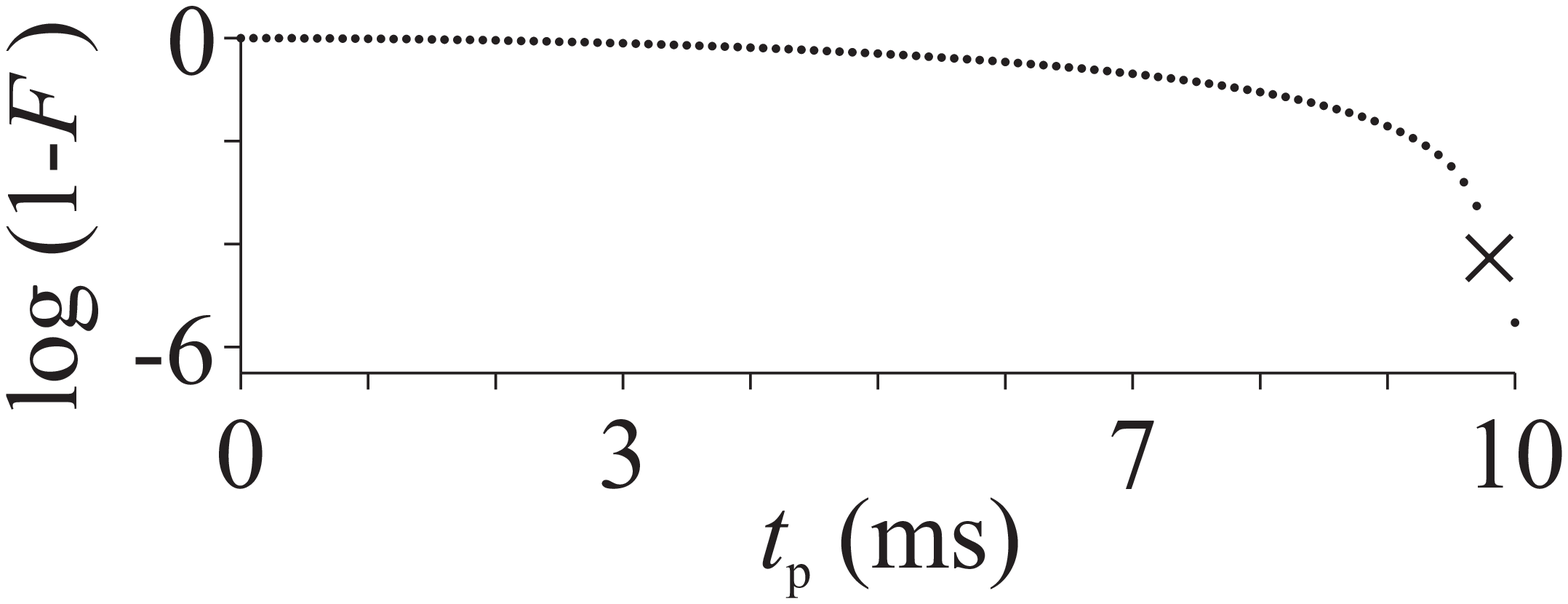}\\
		\\[-3mm]
		\raisebox{0.6cm}{6} &
		\includegraphics[width=0.45\columnwidth]{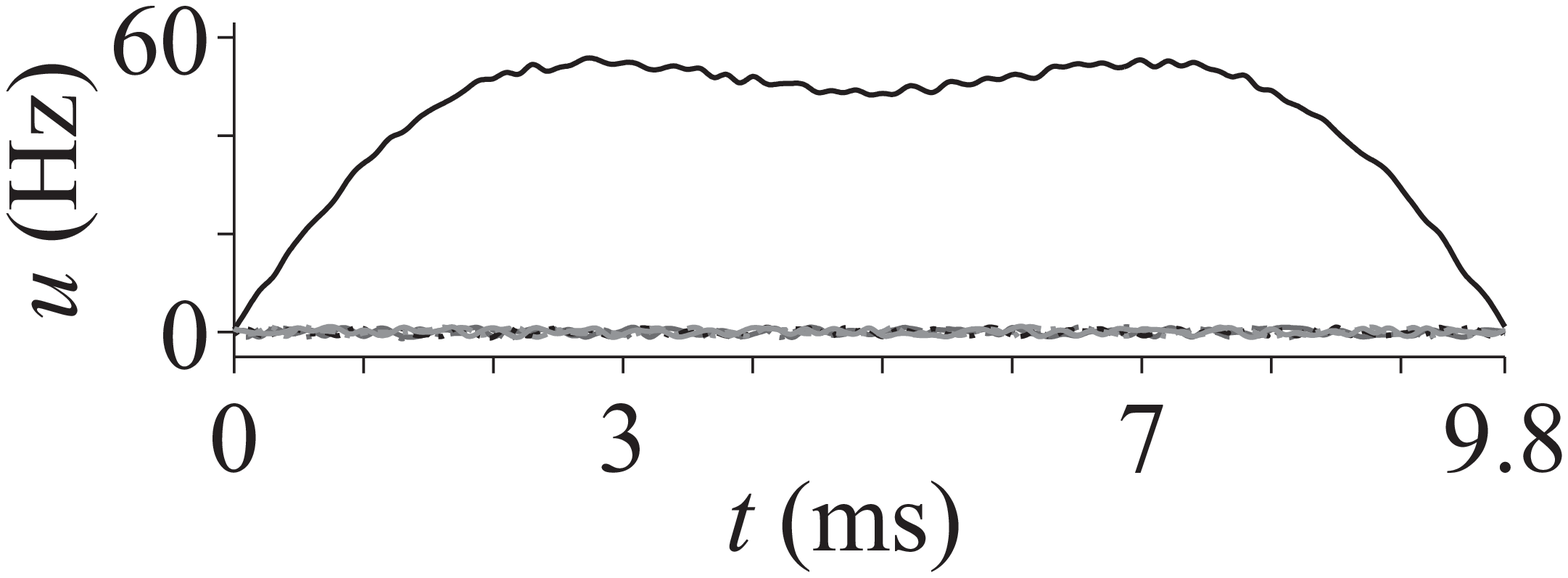} &
		\includegraphics[width=0.45\columnwidth]{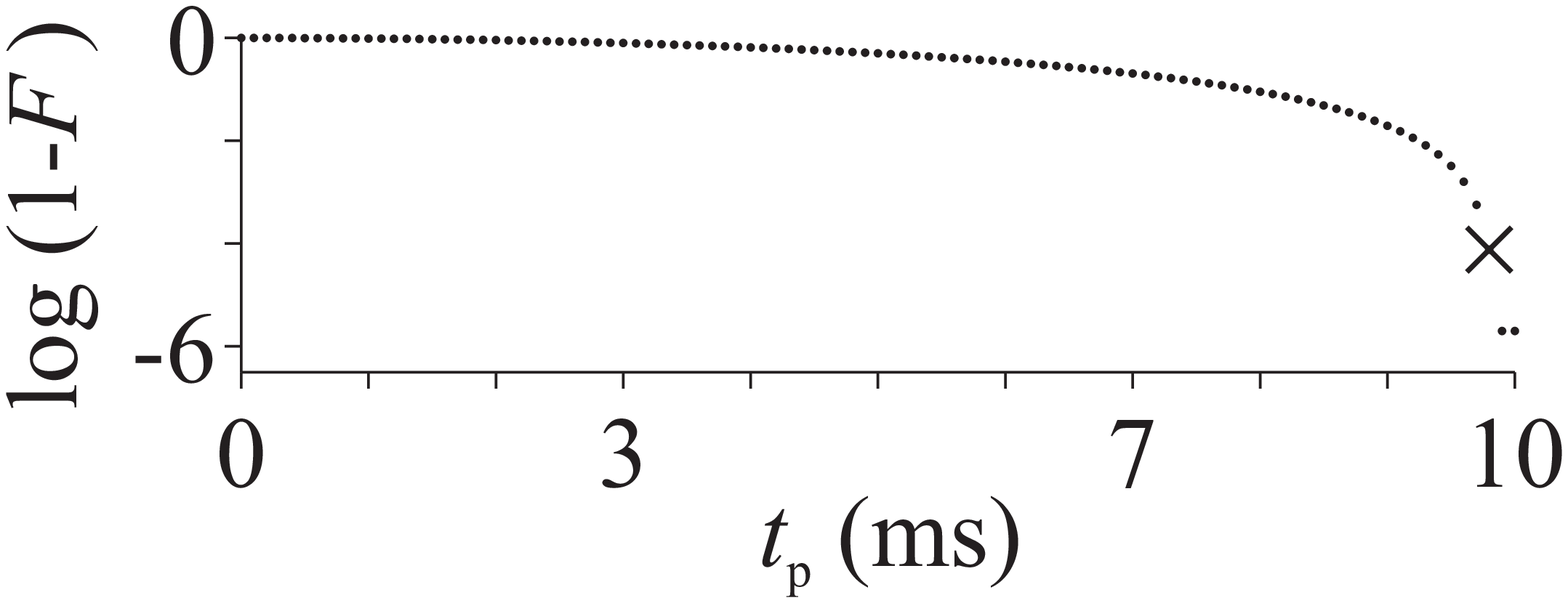}\\\\[-3mm]
		\hline \\[-3.5mm]
		\multicolumn{3}{l}{(b)~$k=1.59$:}\\
		 \raisebox{0.6cm}{1} &
		\includegraphics[width=0.45\columnwidth]{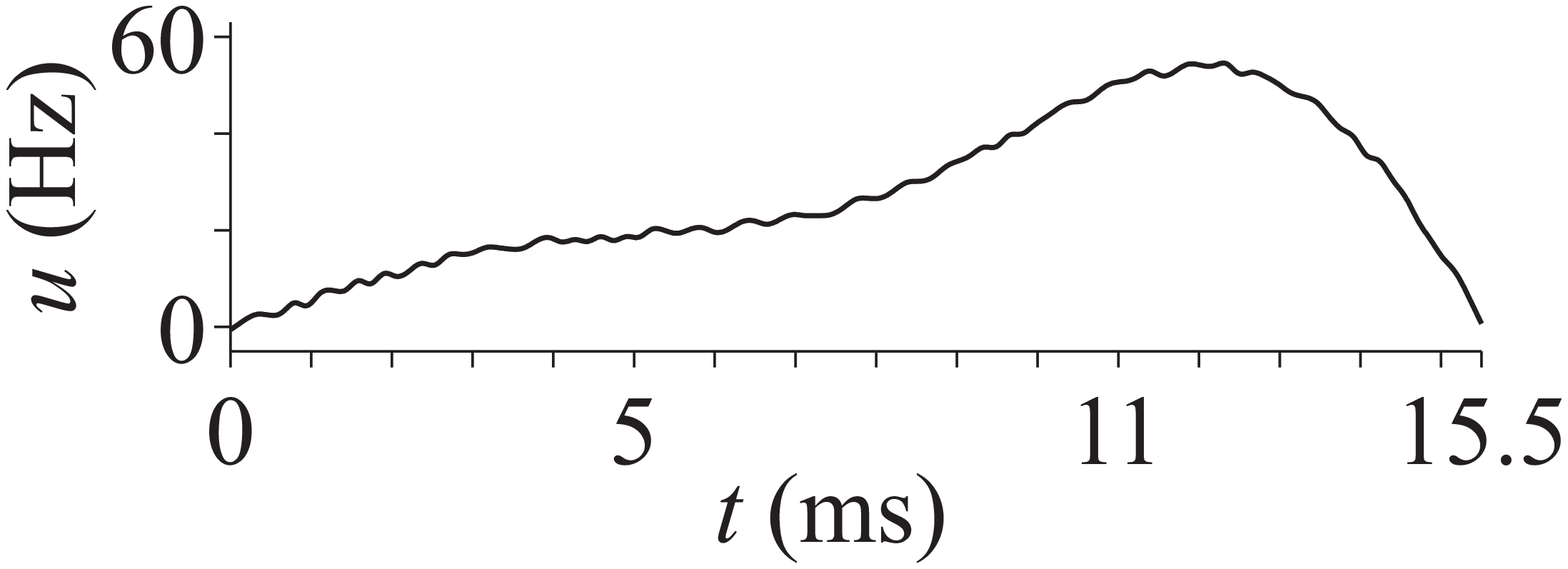} &
		\includegraphics[width=0.45\columnwidth]{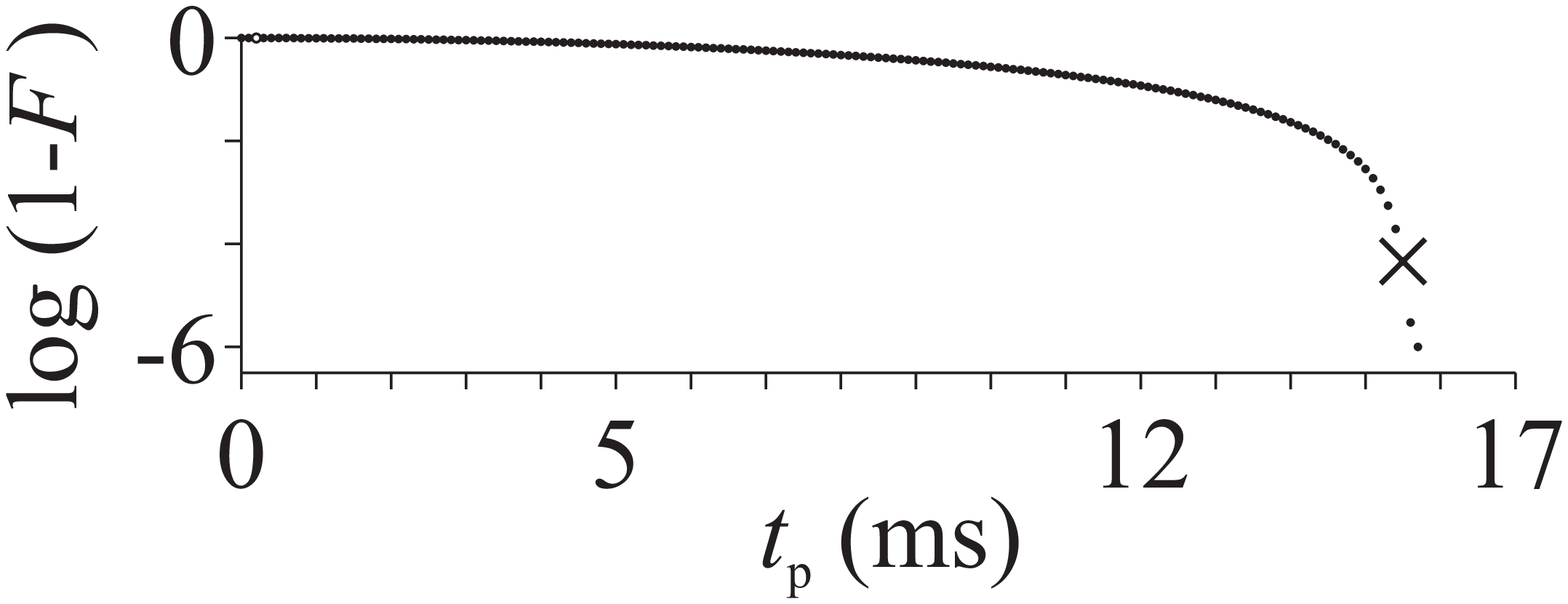}\\
		\\[-3mm]
		\raisebox{0.6cm}{6} &
		\includegraphics[width=0.45\columnwidth]{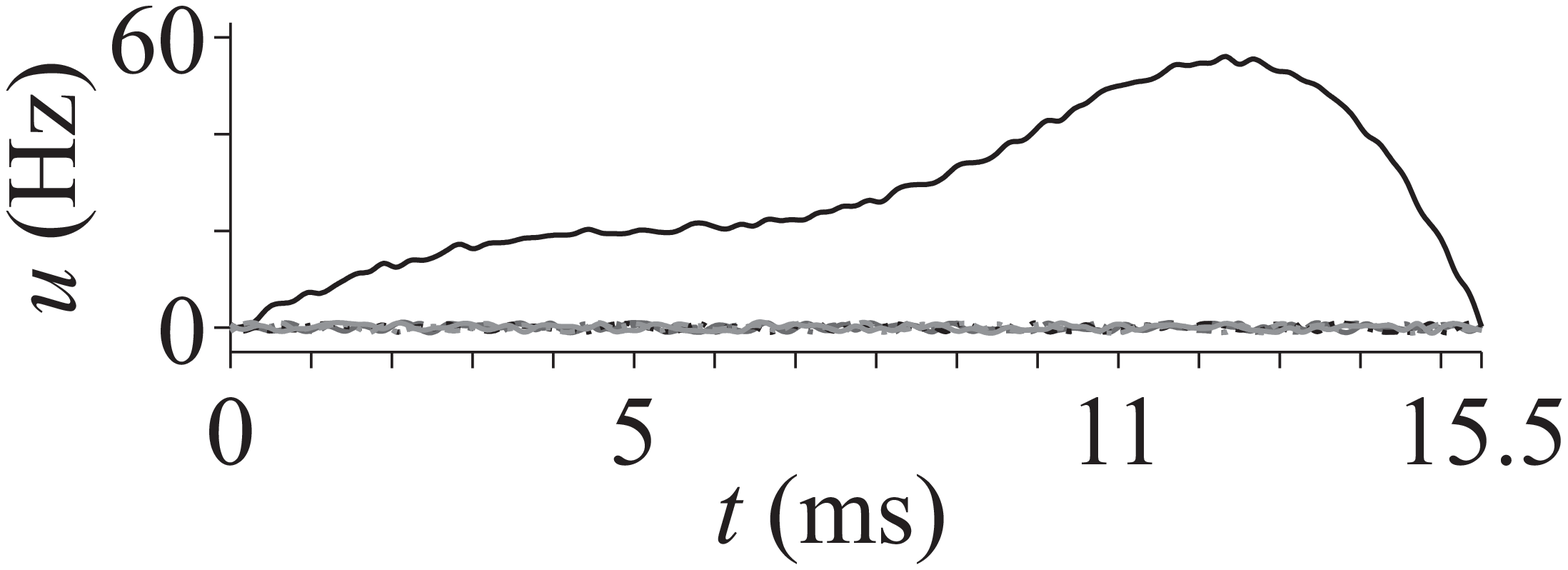} &
		\includegraphics[width=0.45\columnwidth]{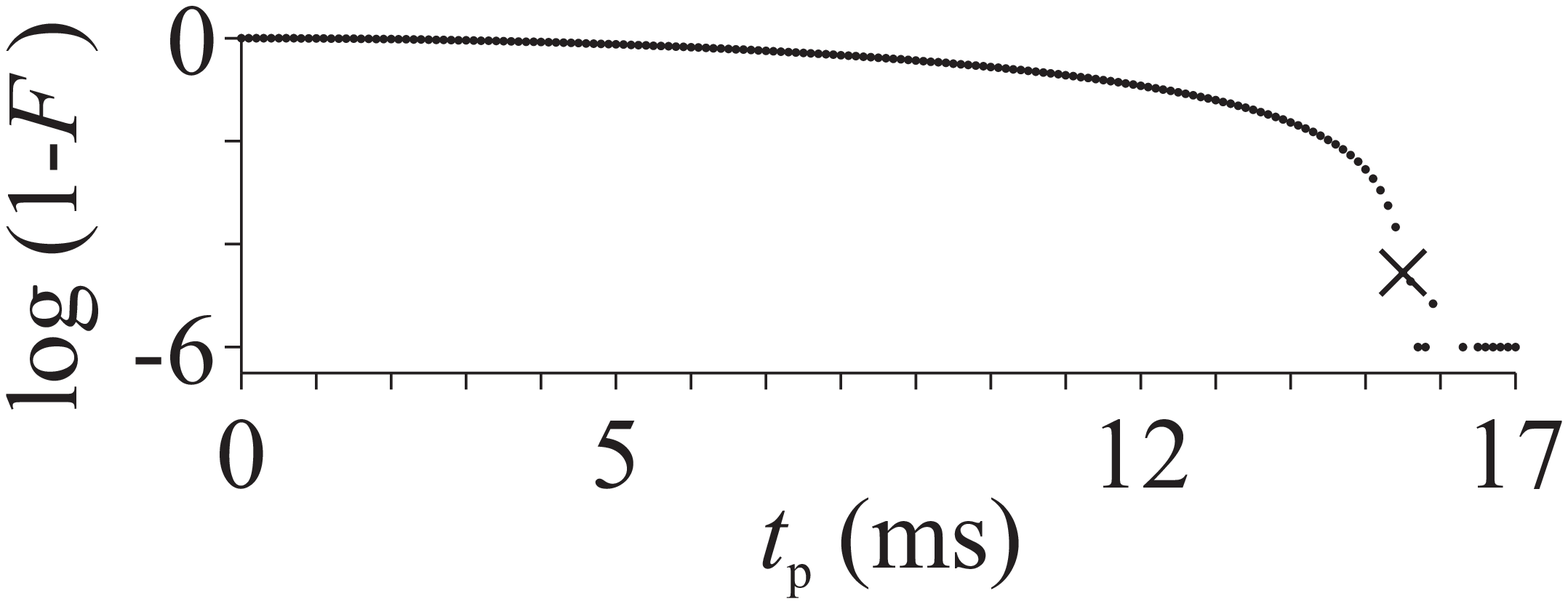}\\
		\end{tabular}
	\end{ruledtabular}
\end{table}

\subsection{Numerical approach\label{sec:lin3spin_numeric}}

We numerically optimize pulse shapes by employing the GRAPE algorithm \cite{KRKSHG05} which
was developed by employing principles of optimal control theory. 
Using a gradient-based optimization we obtain rf controls which 
steer
an initial state 
(or unitary transformation) to a final state (or unitary transformation) while minimizing (e.g.) the duration 
of the pulse. Both the amplitude and the phase of the resulting pulse can be 
smoothly- or noisy-shaped depending on e.g.\
the initial pulse or bounds on the control strength (see, e.g.\ \cite{SKLBBKG06}).

\begin{table}[tb]
 \caption{\label{tab:tab1} For coherence transfers in linear three-spin 
 chains ($k=1$ and $k=1.59$), we give the numerically-optimized times 
 $t_p$ and the fidelities $F$ in the case of one, two, and  six 
 rf controls (see text). The duration $t_p$ is independent of the number 
of controls which suggests that only one rf-control on the middle spin 
is sufficient for the time-optimal coherence transfer.}
	\begin{ruledtabular}
	  \begin{tabular}{c}
		\begin{tabular}{l@{\hspace{4mm}}|@{\hspace{4mm}}r}
     \begin{tabular}{c@{\hspace{3mm}}c@{\hspace{3mm}}c@{\hspace{3mm}}c}
		 $k$ & {$\#u$} & $t_p$~(s) & $F$\\
		 \hline
		 1 & 1 & 0.0098 & 0.9999\\
		 1 & 2 & 0.0098 & 0.9999\\
		 1 & 6 & 0.0098 & 0.9999\\
		 \end{tabular} 
		 &
		 \begin{tabular}{c@{\hspace{3mm}}c@{\hspace{3mm}}c@{\hspace{3mm}}c}
		 $k$  & {$\#u$} & $t_p$~(s) & $F$\\
     \hline
		 1.59 & 1 & 0.0155 & 0.9999\\
		 1.59 & 2 & 0.0155 & 0.9999\\
		 1.59 & 6 & 0.0155 & 0.9999\\
		 \end{tabular}
		\end{tabular}
		\end{tabular}
	\end{ruledtabular}
\end{table}

We treat three different levels of rf controls: First, we use only one rf 
control operating on the second spin along the $y$-direction. Second, we 
use two different rf controls operating on the second spin along both the 
$x$- and $y$-direction. Third, we use a total of six rf controls operating 
on each of the three spins along both the $x$- and $y$-direction. We remark 
that employing rf controls on one spin along both the $x$- and $y$-direction 
gives complete (local) control on that spin. Let $k$ denote the ratio between the 
couplings $J_{23}$ and $J_{12}$. We determine the numerically-optimized pulses 
and plot the logarithmic fidelity $F$ vs. the duration $t_{p}$ of different shaped pulses for
the coupling ratios $k=1$ and $k=1.59$ 
which are motivated by the experimental scenarios of
Fig.~\ref{fig:3_quibit}. 
The numerical results are given in Table~\ref{tab:linear_3spin_k1_shape}:
We show examples 
of shaped pulses with short duration $t_p$ and high fidelity $F \ge 0.9999$.
In addition, we present logarithmic time-optimal (TOP) curves where we plot 
the logarithmic transfer efficiency (i.e.\ $\log(1{-}F)$ where $F$ is the fidelity)
versus the optimal transfer time. 
Comparing the different cases suggests that only one rf control on 
the second spin is sufficient for a time-optimal pulse. For high fidelities
($F \ge 0.9999$), the durations of the analytic and numerically-optimized 
pulses are identical (in the given accuracy) while the pulse forms differ. 
In Table~\ref{tab:tab1}, 
we compare the duration of pulses on linear three-spin systems 
for different values of $k$.

\begin{conj}\label{conj1}
Consider a linear three-spin chain with local controls on each spin.
One can time-optimally transfer
coherence from $I_{1x}$ to $4 I_{1y} I_{2y} I_{3z}$
using only one control on the second spin along the $y$-direction.
In addition, the analytical pulses of Refs.~\cite{YGK07,YZK08}
are time-optimal in the case of linear three-spin chains even if one allows arbitrary local controls.
\end{conj}

\section{Linear four-spin chains: analytic and numerical approaches\label{sec:3_4_spin}}

\subsection{Analytic approach\label{sec:3_4_spin_theory}}
\begin{figure}[tb]
	\includegraphics[width=0.7\columnwidth]{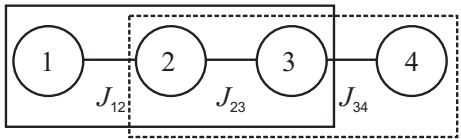}
	\caption{A linear four-spin chain has only direct couplings 
	$J_{12}$, $J_{23}$, and $J_{34}$ between
	neighboring spins. We split the corresponding four-spin chain control 
	problem into two subproblems for three-spin chains.}
	\label{fig:ising4spinsytem}
\end{figure}

In this section, we consider linear spin chains with four spins.
We follow Sect.~IV of Ref.~\cite{YZK08} (see also \cite{YGK07})
and split the control problem for four spins into two subproblems
for three spins (see Fig.~\ref{fig:ising4spinsytem}):
The first subproblem is given on the first three spins
by the time-optimal transfer 
from $(1,0,0)^T$ to $(0,\cos\gamma,\sin\gamma)^T$, where
we are again using the coordinates $(r_1,r_2,r_3)^T$ of Sect.~\ref{sec:lin3spin_theory}.
Then, we apply certain (arbitrarily fast) hard pulses which
can be easily determined by numerical methods.
The second subproblem is given on the last three spins
by the time-optimal 
transfer 
from $(\cos\gamma,\sin\gamma,0)^T$ to $(0,0,1)^T$.
In addition, we have to simultaneously search
for the value of $\gamma$ which minimizes the pulse duration.
This approach might not lead to time-optimal controls but simplifies the
control problem significantly. 

\begin{figure}[tb]
	\includegraphics[width=0.49\columnwidth]{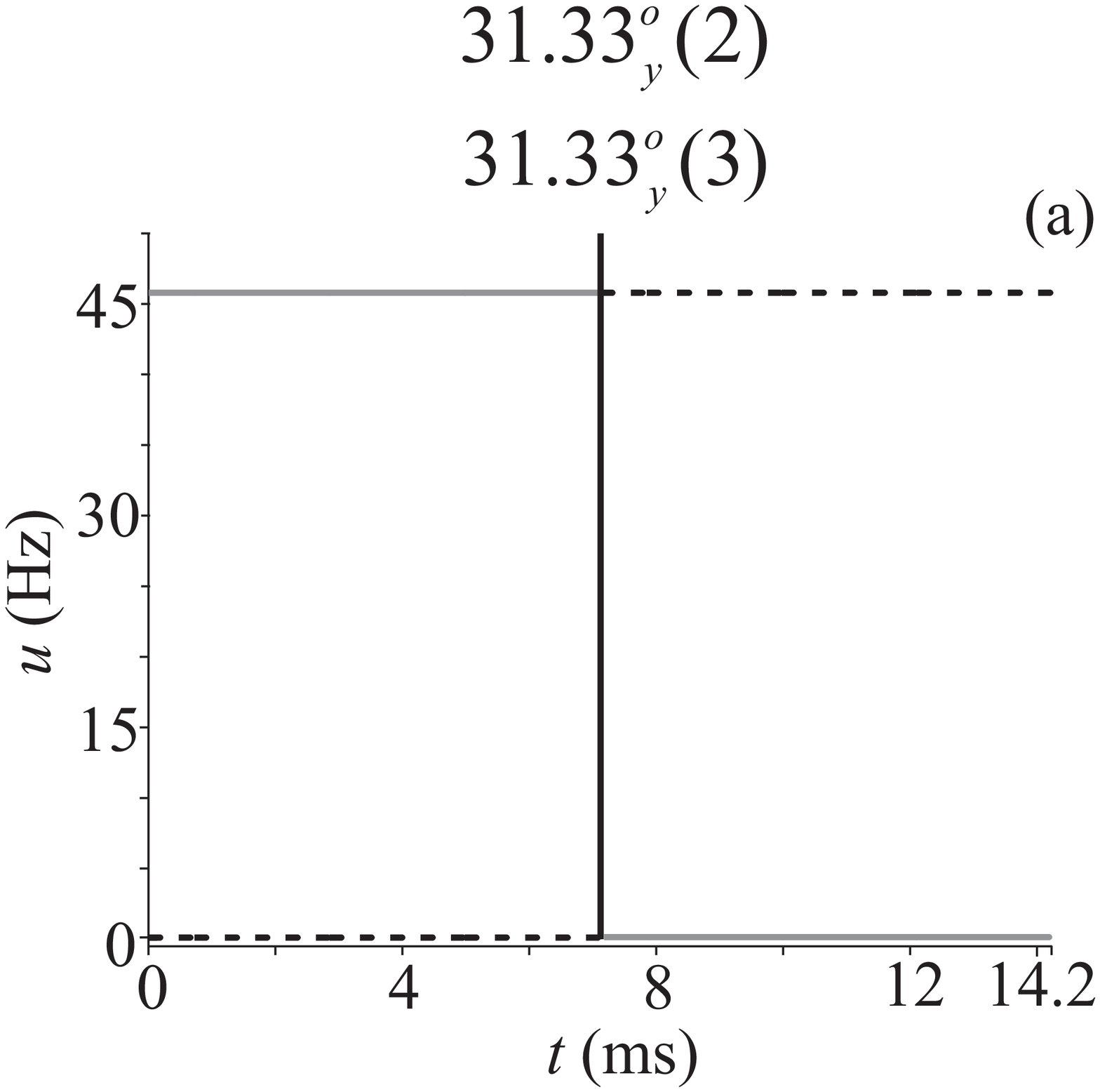} \includegraphics[width=0.49\columnwidth]{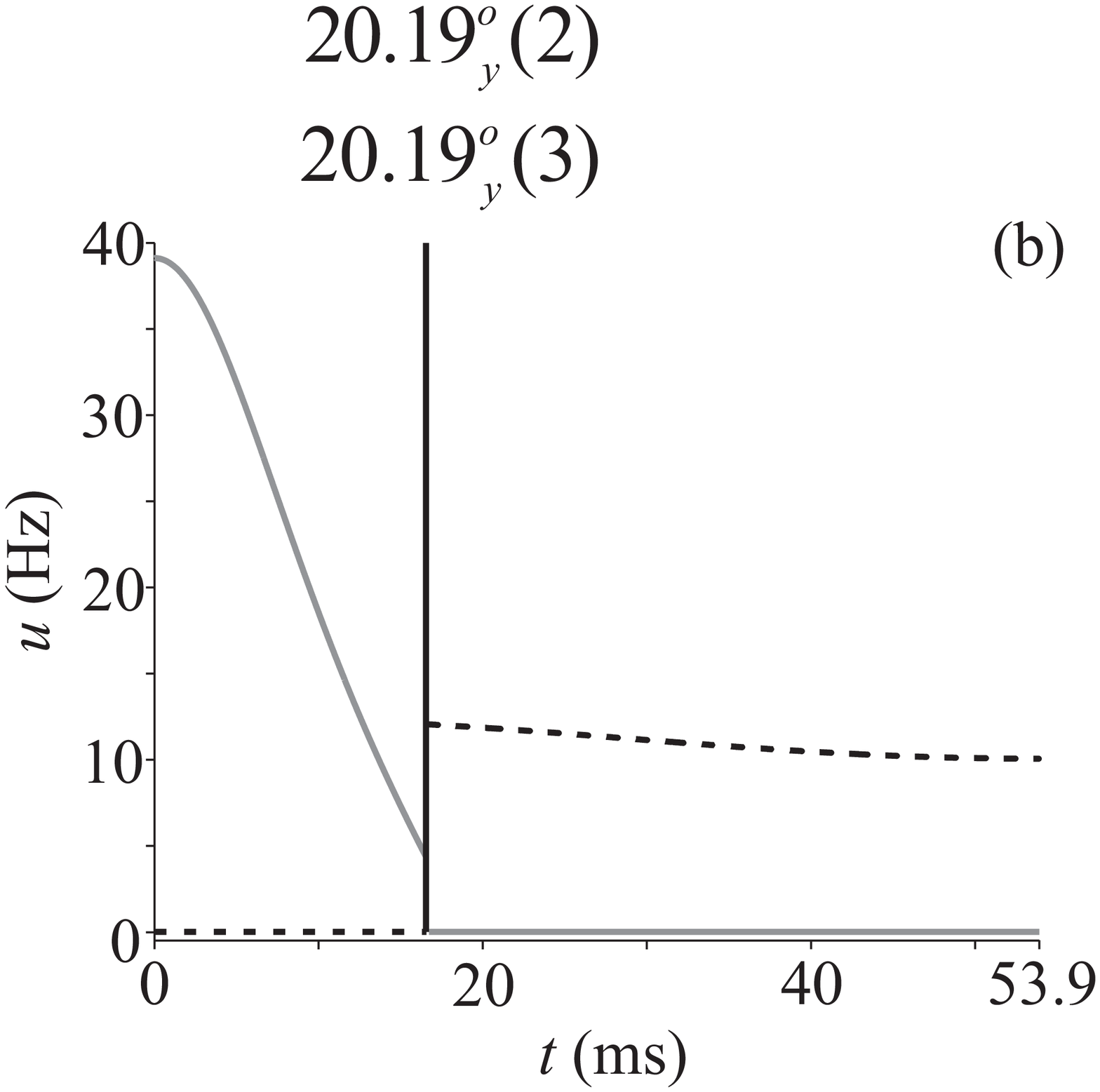}
	\caption{Analytic pulses for linear four-spin chains are given in the cases of (a) $k_1=k_2=1$ as well as (b) $k_1=2.38$ and $k_2=0.94$. The pulses on the second and third spin along the $y$-direction are given, respectively, as solid and dashed line.
	The corresponding two hard pulses on the second and third spin are depicted by a
	vertical line where the flip angles are given above. 
	The hard pulses in the left figure can be implemented
	by applying a pulse of $5000~\text{Hz}$ for $17.40$ microseconds.
	The hard pulses in the right figure can be implemented
	by applying a pulse of $5000~\text{Hz}$ for $11.21$ microseconds.}
	\label{fig:ising4spin_algebraic_shape}
\end{figure}

\begin{figure}[tb]
	\includegraphics[width=0.5\columnwidth]{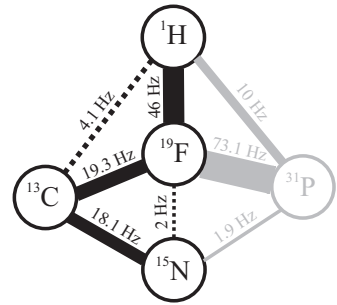}
	\caption{The topology of the molecule
${{}^{13}}\mathrm{C}^{\mathrm{O}}$-${{}^{15}}\mathrm{N}$-diethyl-(dimethylcarbonyl)fluoromethylphosphonate (see \cite{pomplun:2010,MPBZEFG11}) results in coupling ratios $k_1=2.38$ and $k_2=0.94$.
	Compare to Fig.~\ref{fig:3_quibit}. 
	}
	\label{fig:5_quibit}
\end{figure}

The optimization of the considered subproblems can be reduced to time-optimal transfers
from $(\cos\alpha,\sin\alpha,0)^T$ to $(0,\cos\beta,\sin\beta)^T$ for $\alpha,\beta \in
[0,\pi/2]$  generalizing the transfer of Sect.~\ref{sec:lin3spin_theory}
from $(1,0,0)^T$ to $(0,0,1)^T$.
Using methods of Ref.~\cite{YZK08} we can find the 
optimal controls for the transfers using combined one-dimensional 
searches for the optimal initial values $\theta(0)$ and $\dot{\theta}(0)$ of Eq.~\eqref{theta_eq}.
Both $\theta(0)$ and $\dot{\theta}(0)$ are undetermined
but related by $\dot{\theta}(0)=\sin[\theta(0)] \cot\alpha$ for the case of
$(r_1(0),r_2(0),r_3(0))^T=(\cos\alpha,\sin\alpha,0)^T$.
The corresponding (semi-)analytic pulses 
are shown in Fig.~\ref{fig:ising4spin_algebraic_shape}.
The values are motivated by the experimental system given in
Fig.~\ref{fig:5_quibit}.

\subsection{Numerical approach\label{sec:3_4_spin_numeric}}

\begin{table}[tb]
\caption{\label{tab:linear_4spin_k1_shape}%
For linear four-spin chains with coupling ratios (a) 
$k_1=1$ and $k_2=1$ as well as (b) $k_1=2.38$ and $k_2=0.94$, 
the rf controls on the second and third spin along the $y$-axis 
are plotted using solid-black and solid-gray lines, respectively. 
Other rf controls are plotted using a dashed-black line or in 
shades of gray. Compare to Table~\ref{tab:linear_3spin_k1_shape}.}
	\begin{ruledtabular}
		\begin{tabular}[t]{c  c  c}
		$\#{}u$ & pulse shape  & logarithmic TOP curve\\
		\hline \\[-3mm]
		\multicolumn{3}{l}{(a)~$k_1=1$ and $k_2=1$:}\\
		 \raisebox{0.6cm}{2} &
		\includegraphics[width=0.45\columnwidth]{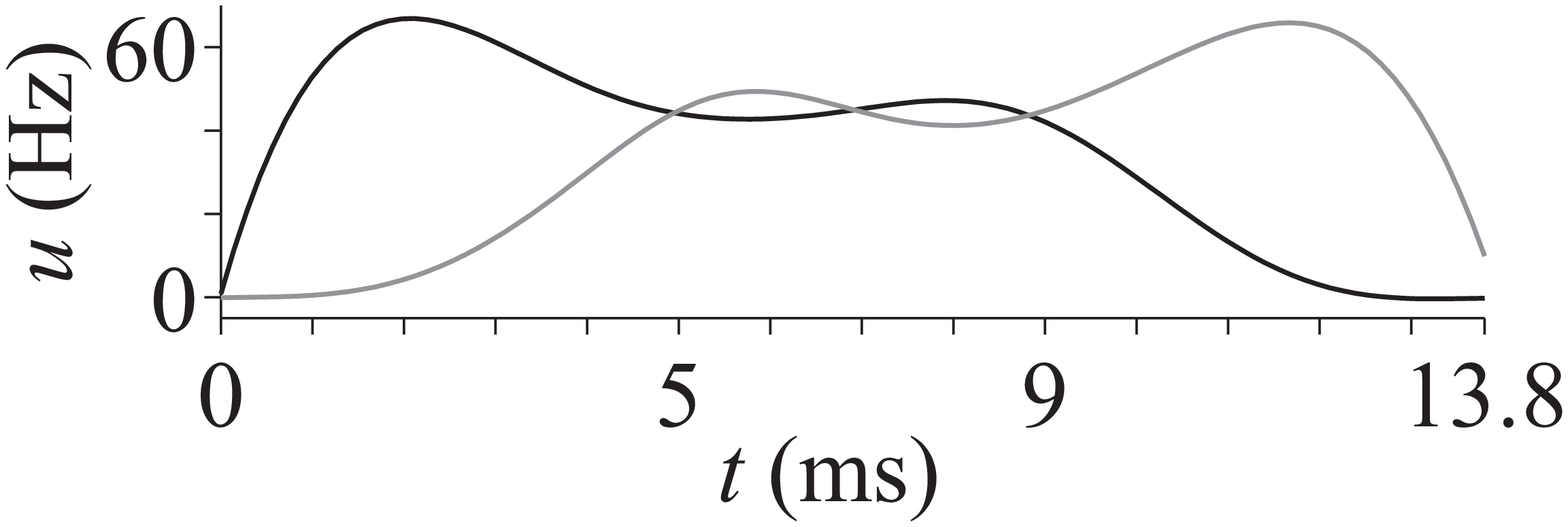} &
		\includegraphics[width=0.45\columnwidth]{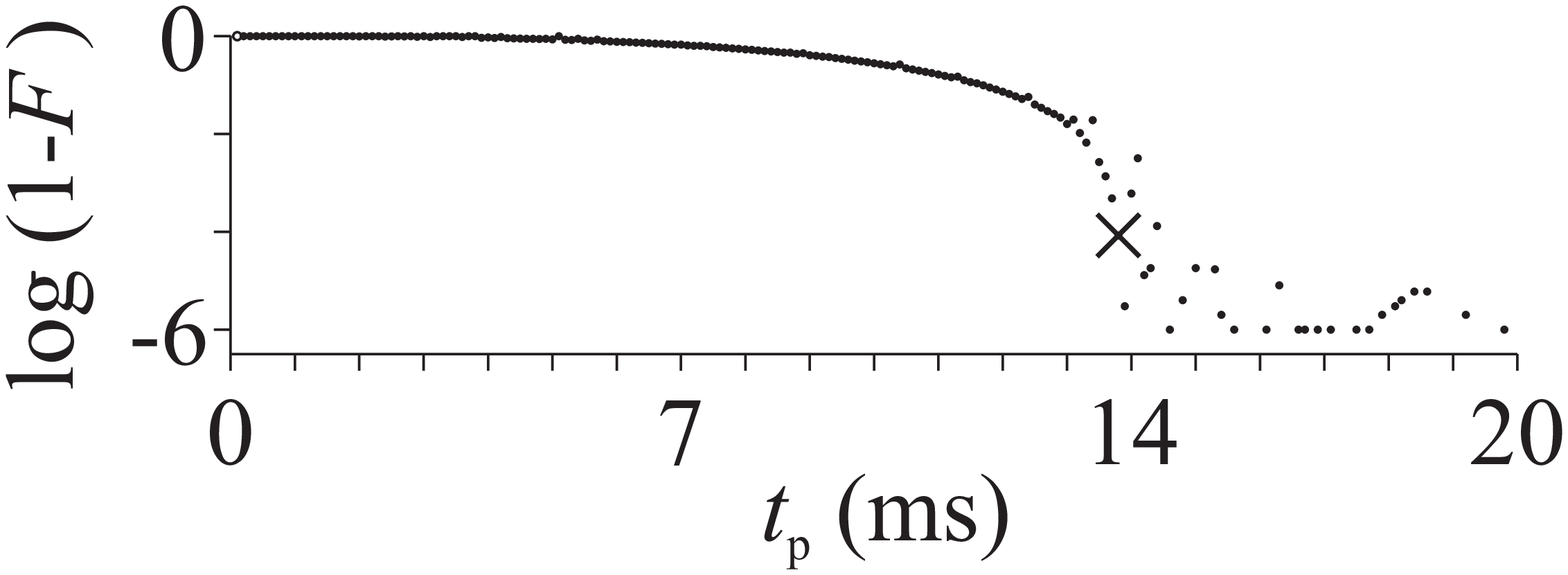}\\
		\\[-3mm]
		\raisebox{0.6cm}{8} &
		\includegraphics[width=0.45\columnwidth]{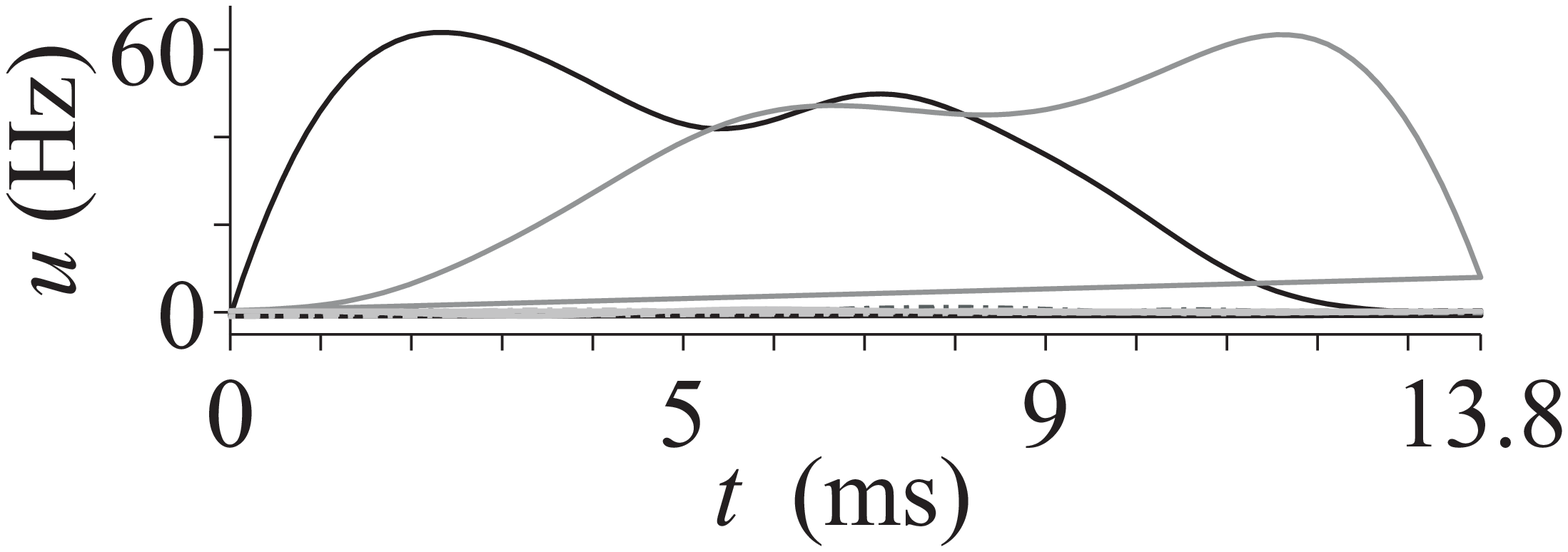} &
		\includegraphics[width=0.45\columnwidth]{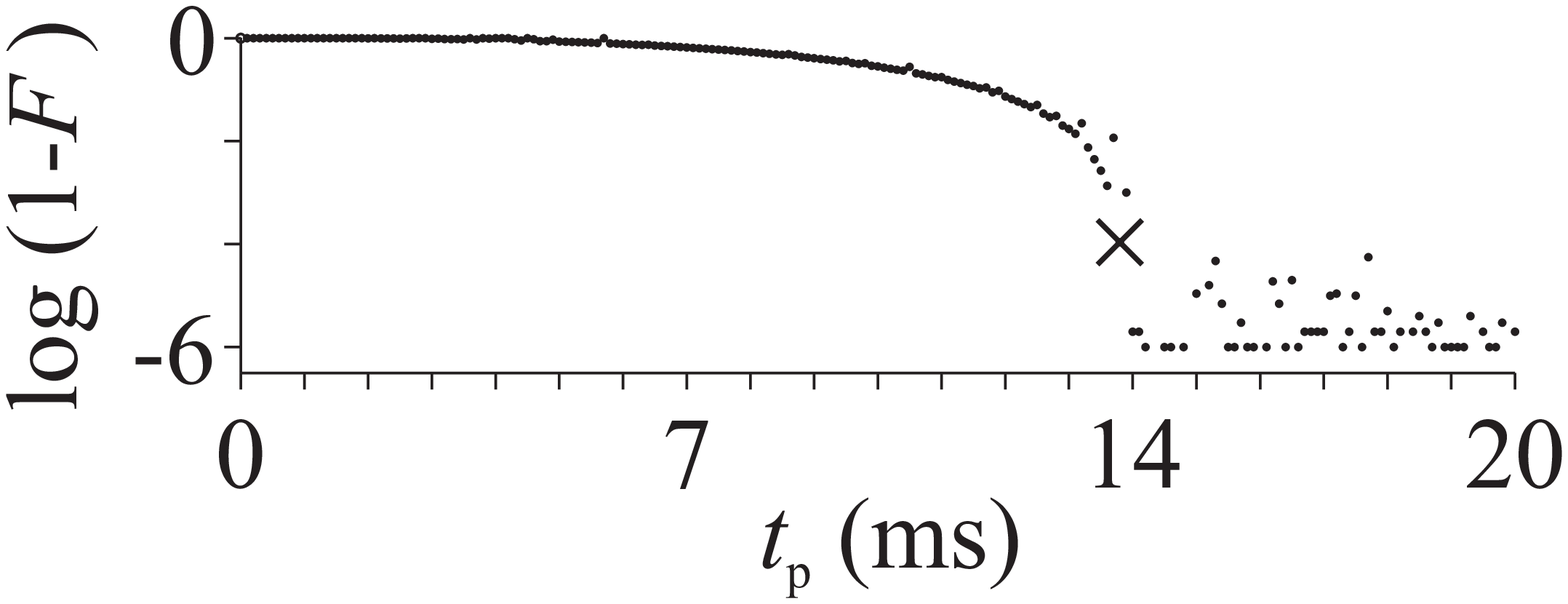}\\\\[-3mm]
		\hline \\[-3.5mm]
		\multicolumn{3}{l}{(b)~$k_1=2.38$ and $k_2=0.94$:}\\
		 \raisebox{0.6cm}{2} &
		\includegraphics[width=0.45\columnwidth]{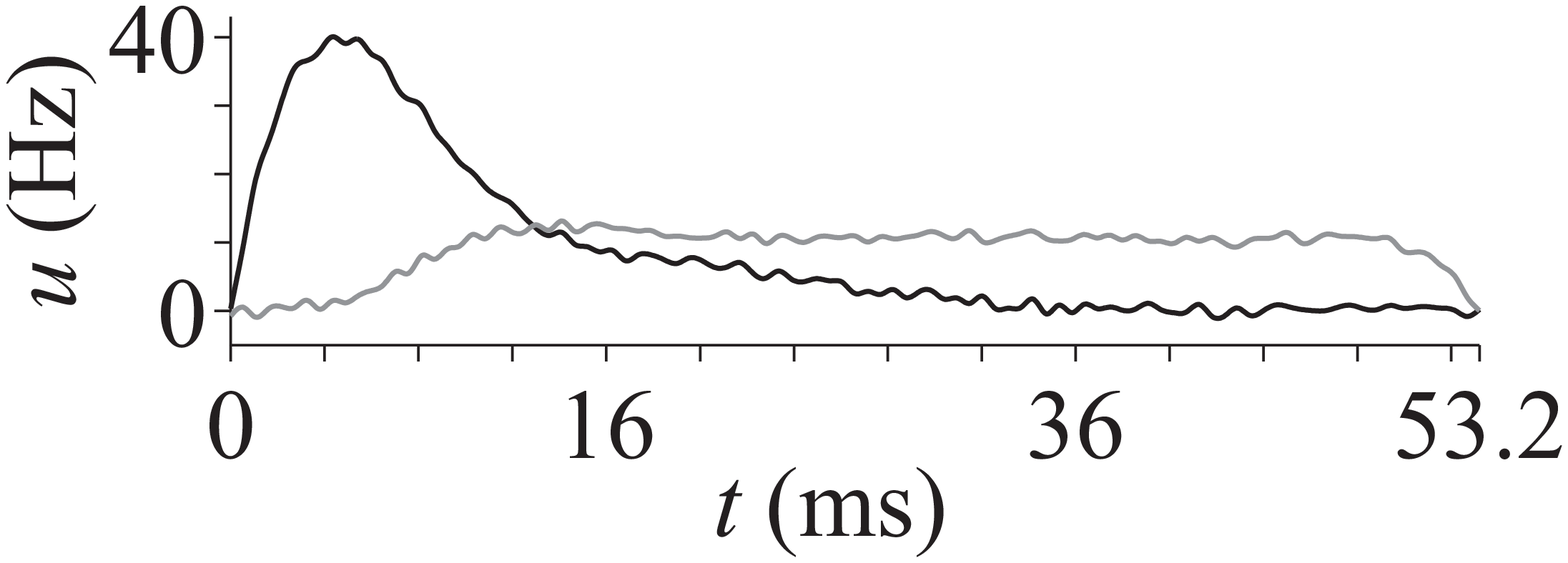} &
		\includegraphics[width=0.45\columnwidth]{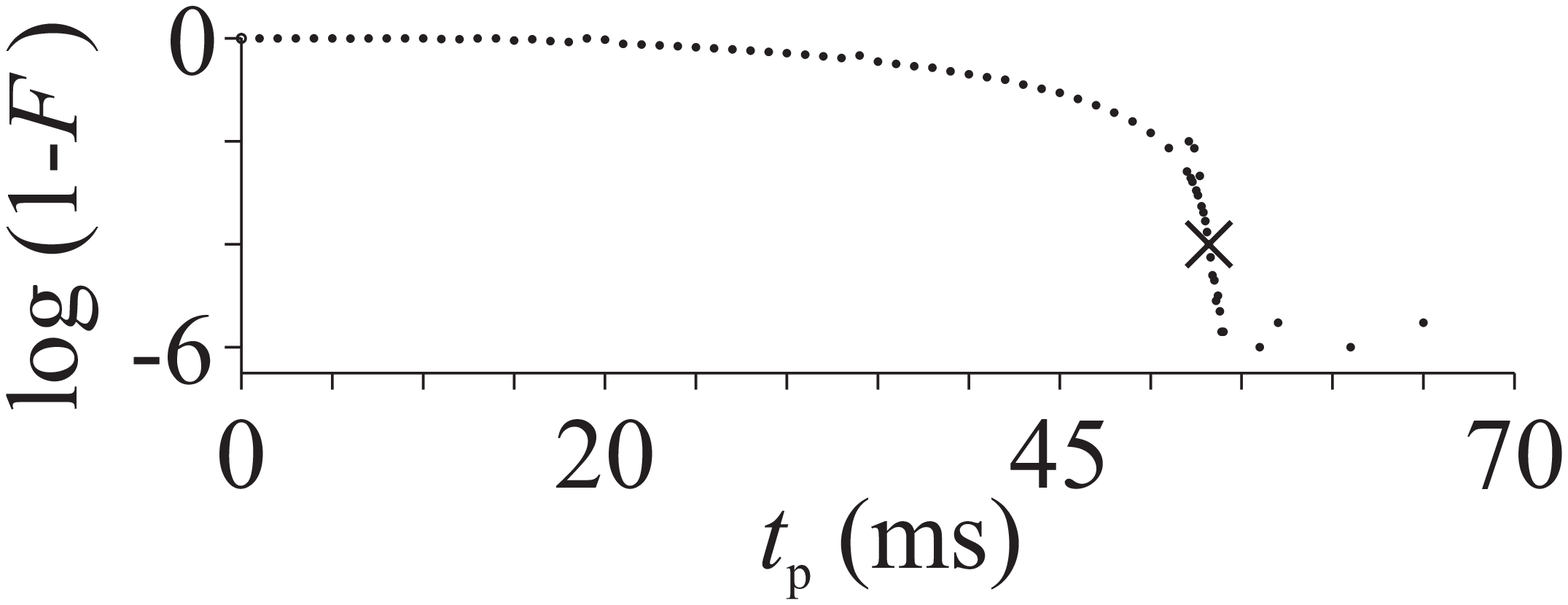}\\
		\\[-3mm]
		\raisebox{0.6cm}{8} &
		\includegraphics[width=0.45\columnwidth]{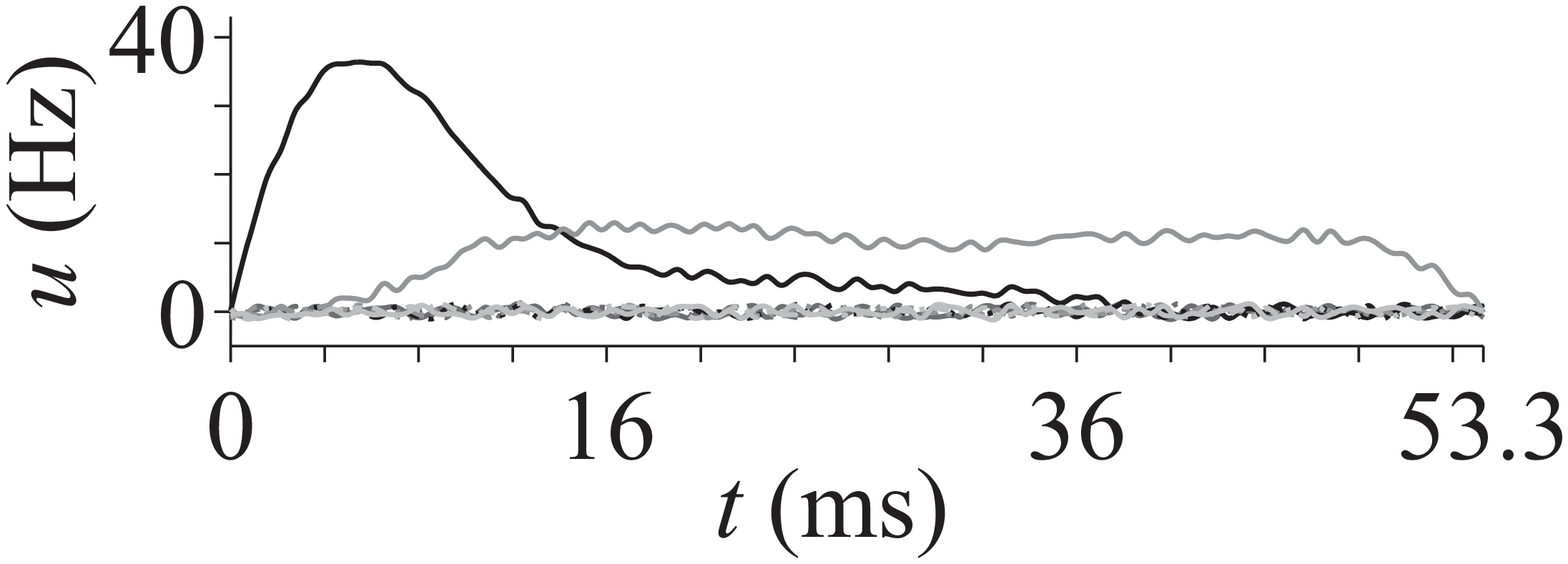} &
		\includegraphics[width=0.45\columnwidth]{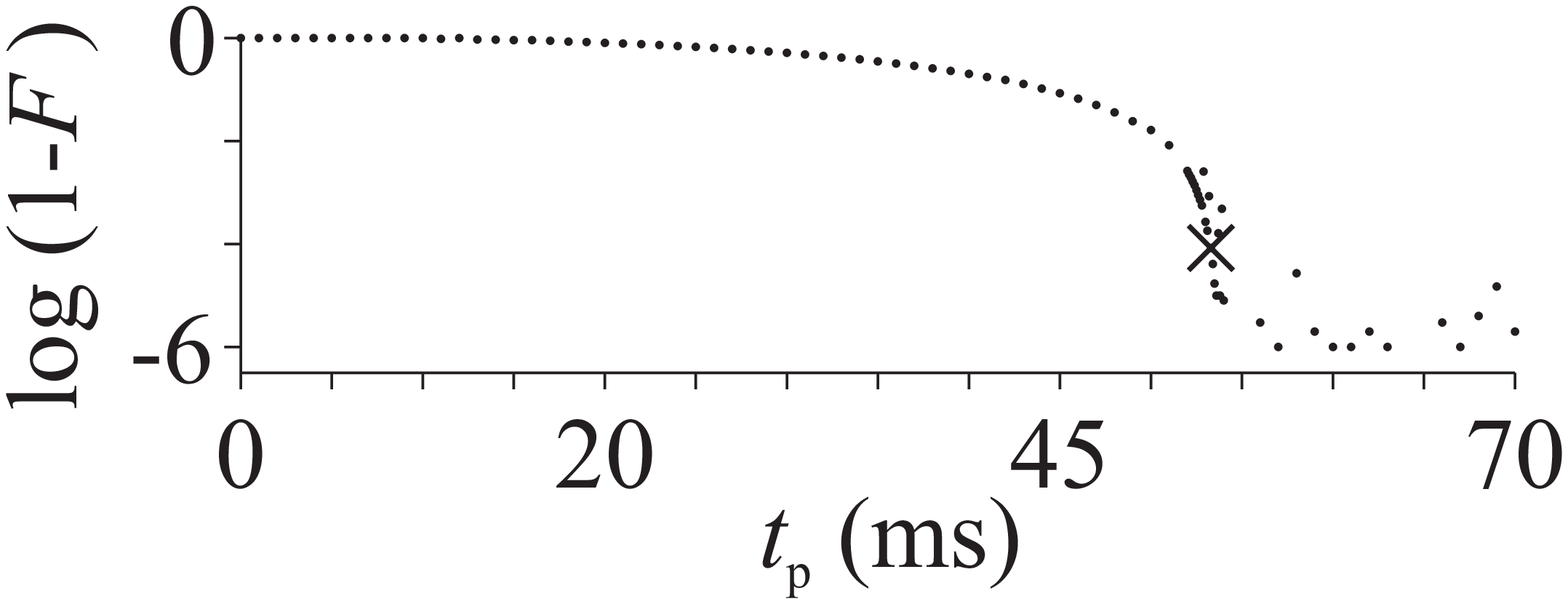}\\
		\end{tabular}
	\end{ruledtabular}
\end{table}

Motivated by the analytical approach, we numerically treat the control 
problem on four spins with two cases of coupling ratios (a) $k_1=1$ 
and $k_2=1$ ($J_{12}=J_{23}=J_{34}=88.05$~Hz), and (b) $k_1=2.38
\approx J_{12}/J_{23}$ and 
$k_2=0.94\approx J_{34}/J_{23}$ (refer to Fig.~\ref{fig:5_quibit} for the coupling values). The coherence transfer is 
numerically optimized considering the following three levels of rf 
controls: First, we use only two different rf controls (one on each spin) 
operating on the second and third spin along the $y$-direction. Second, we use 
a total of four different rf controls (two on each spin) operating on 
the second and third spin along both 
the $x$- and $y$-direction. Third, 
we use a total of eight different rf controls (two on each spin) operating on each 
of the four spins along both the $x$- and $y$-direction. The pulse shapes 
and the logarithmic TOP curves corresponding to two and eight rf controls (see 
Table~\ref{tab:linear_4spin_k1_shape}) indicate that we do not gain a higher 
fidelity or a shorter duration from using more than the two controls (Table~\ref{tab:tab_four}). This is 
consistent with the analytic results, but the numerically-optimized pulses 
appear to be a little shorter than the analytical ones (cp.\ Fig.~\ref{fig:ising4spin_algebraic_shape}).

\begin{conj}\label{conj2}
Consider a linear four-spin chain with local controls on each spin.
One can time-optimally transfer
coherence from $I_{1x}$ to $8 I_{1y} I_{2y} I_{3y} I_{4z}$
using only two controls along the $y$-direction
which operate on the second and third spin, respectively.
\end{conj}

\begin{table}[tb]
 \caption{\label{tab:tab_four} For coherence transfers in linear four-spin 
 chains ($k_1=k_2=1$ as well as $k_1=2.38$ and $k_2=0.94$), we give the numerically-optimized times 
 $t_p$ and the fidelities $F$ in the case of two, four, and  eight 
 rf-controls. The duration $t_p$ is independent of the number 
of controls which suggests that only two rf-control on the second and third spin
along the $y$-axis 
is sufficient for the time-optimal coherence transfer.}
	\begin{ruledtabular}
	  \begin{tabular}{c}
		\begin{tabular}{l@{\hspace{2mm}}|@{\hspace{2mm}}r}
     \begin{tabular}{c@{\hspace{2mm}}c@{\hspace{2mm}}c@{\hspace{2mm}}c@{\hspace{2mm}}c}
		 $k_1$ & $k_2$ & {$\#u$} & $t_p$~(s) & $F$\\
		 \hline
		 1 & 1 & 2 & 0.138 & 0.9999\\
		 1 & 1 & 4 & 0.138 & 0.9999\\ 
		 1 & 1 & 8 & 0.138 & 0.9999\\
		 \end{tabular} 
		 &
		 \begin{tabular}{c@{\hspace{2mm}}c@{\hspace{2mm}}c@{\hspace{2mm}}c@{\hspace{2mm}}c}
		 $k_1$ & $k_2$ & {$\#u$} & $t_p$~(s) & $F$\\
     \hline
     2.38 & 0.94 & 2 & 0.532 & 0.9999\\
     2.38 & 0.94 & 4 & 0.532 & 0.9999\\
     2.38 & 0.94 & 8 & 0.533 & 0.9999\\
		 \end{tabular}
		\end{tabular}
		\end{tabular}	
	\end{ruledtabular}
\end{table}

\section{More generally-coupled spin systems of three and four 
spins\label{sec:3_4_spin_loopedspin}}

In more generally-coupled spin systems, indirect couplings can strongly 
impede or enhance the coherence transfer. In this section we present 
detailed numerical optimizations and compare them to the case of linear 
spin chains.

\subsection{Three-spin system\label{sec:3_4_spin_loopedspin_3spin}}

\begin{figure}[tb!h]
	\includegraphics[width=0.96\columnwidth]{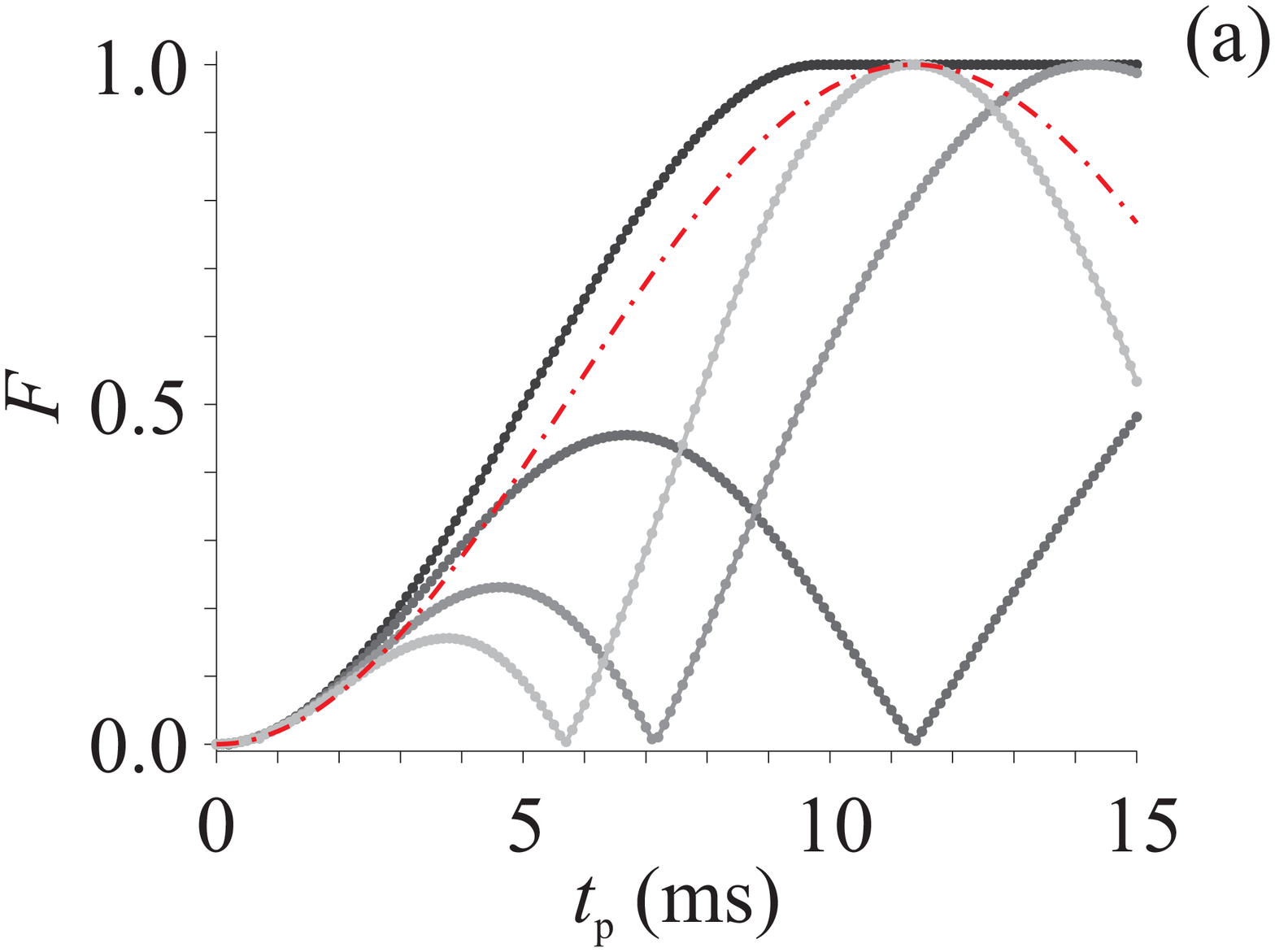}\\[+3mm]
	\includegraphics[width=0.96\columnwidth]{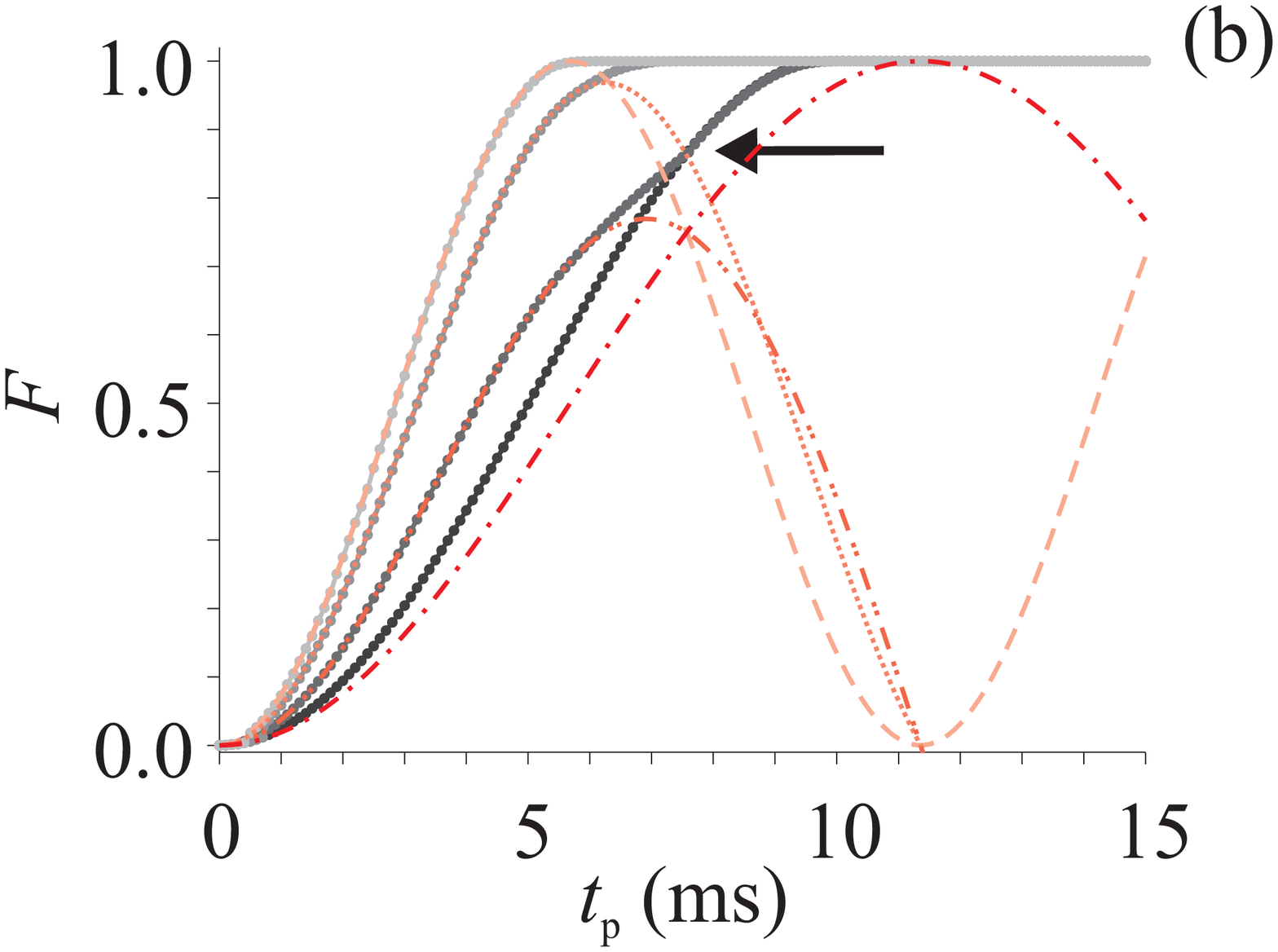}\\[+3mm]
	\includegraphics[width=1\columnwidth]{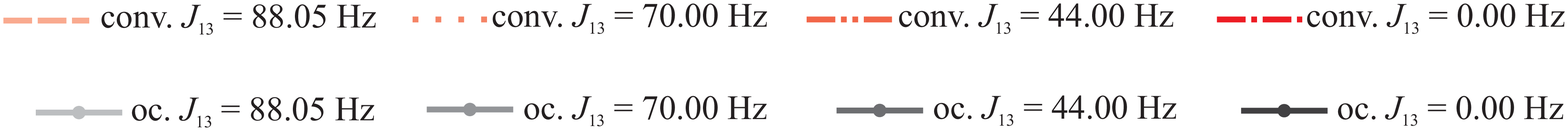}
	\caption{(Color online) We compare numerically-optimized TOP curves (in shades of gray) for three-spin systems with one [Subfigure~(a)] and six [Subfigure~(b)] rf controls keeping $J_{12}=J_{23}=88.05~\text{Hz}$ constant while varying $J_{13}$. At the same time we compare their performance with conventional pulse sequences 
	 (in shades of red) where both of the couplings $J_{12}$ and $J_{13}$ evolve simultaneously. The limiting case of $J_{13}=0$ corresponds to the conventional pulse sequence of Sec.~\ref{model}.
	Using all six rf controls [Subfigure~(b)] we can see higher fidelities $F$  for smaller times and larger $J_{13}$ compared to 
	the case of only one rf control [Subfigure~(a)]. A black arrow denotes 	
	 where the numerically-optimized TOP curves for $J_{13}=44~\text{Hz}$ and $J_{13}=0~\text{Hz}$ merge.	
	}
	\label{fig:looped_3spin_J13_test}
\end{figure}

Along the lines of Section~\ref{sec:lin3spin_numeric}, we numerically 
optimize pulses for more generally-coupled three-spin systems keeping 
$J_{12}=J_{23}=88.05~\text{Hz}$  constant while 
varying the additional coupling strength $J_{13}$. By comparing the TOP 
curves for different values of $J_{13}$, we conclude that for a larger
coupling strength $J_{13}$ the fidelity of the coherence transfer is 
smaller in the cases of one [Fig.~\ref{fig:looped_3spin_J13_test}(a)]
and two (results are not shown)
rf controls on the second spin. (As in Section~\ref{sec:lin3spin_numeric}, 
we obtain shorter pulse sequences as compared to the conventional pulse sequence
for $J_{13}=0$.)
However, using the 
rf controls on each of the three spins allows for a coherence transfer 
with higher fidelity while keeping the pulses short 
[Fig.~\ref{fig:looped_3spin_J13_test}(b)]. Table~\ref{tab:loop_3spin_k1_shape} 
shows examples of shaped pulses and the corresponding logarithmic TOP curves 
for the coupling ratios $k=1$ and $k=1.59$. The coupling strengths are 
taken from the spin systems shown in Fig.~\ref{fig:3_quibit}.
Detailed values are given in Table~\ref{tab:tab2}.

\begin{table}[tb]
\caption{\label{tab:loop_3spin_k1_shape}%
Numerical results for more generally-coupled three-spin systems 
with (a) coupling ratio $k=1$ (i.e.\ $J_{12}=J_{23}=88.05~\text{Hz}$) and additional coupling 
$J_{13}=2.935~\text{Hz}$ as well as (b) coupling ratio $k=1.59$ 
(i.e.\ $J_{12}=73.1~\text{Hz}$ and $J_{23}=46~\text{Hz}$)
and additional coupling $J_{13}=10.0~\text{Hz}$. Compare to
Table~\ref{tab:linear_3spin_k1_shape}.}
	\begin{ruledtabular}
		\begin{tabular}[t]{c  c  c  }
		$\#{}u$ & pulse shape & logarithmic TOP curve\\
		\hline \\[-3mm]
		\multicolumn{3}{l}{(a)~$k=1$:}\\
		 \raisebox{0.6cm}{1} &
		\includegraphics[width=0.45\columnwidth]{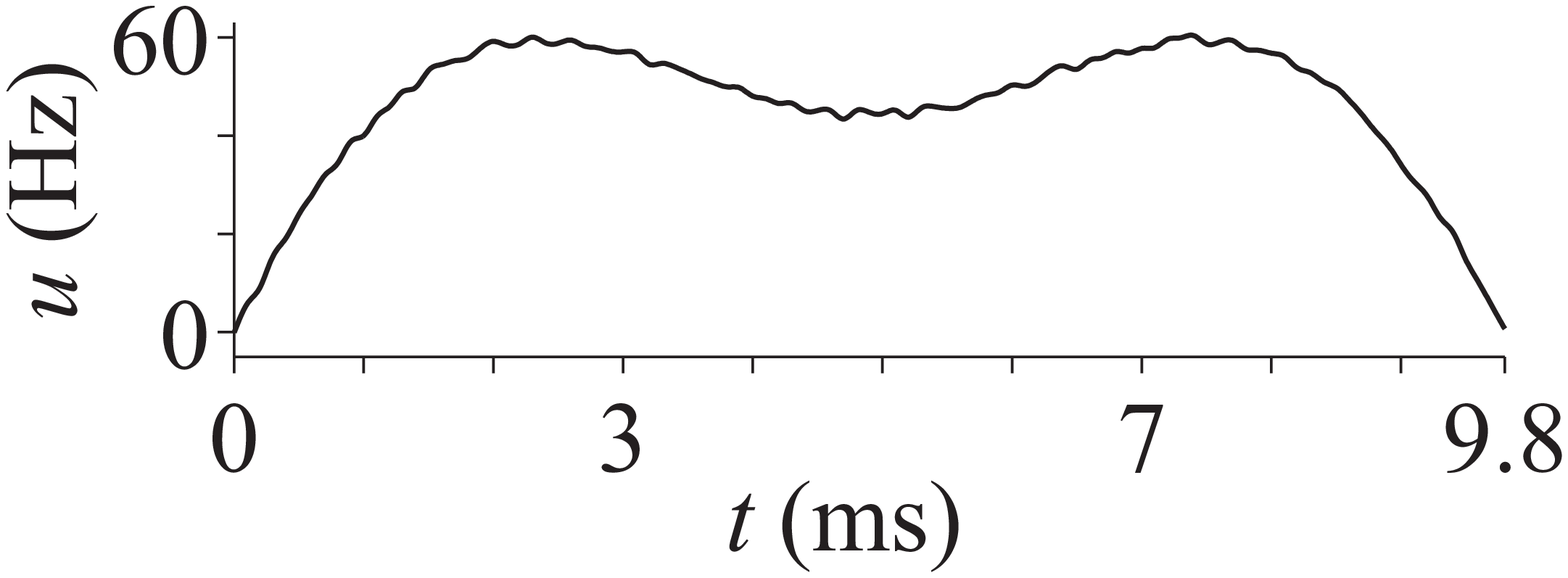} &
		\includegraphics[width=0.45\columnwidth]{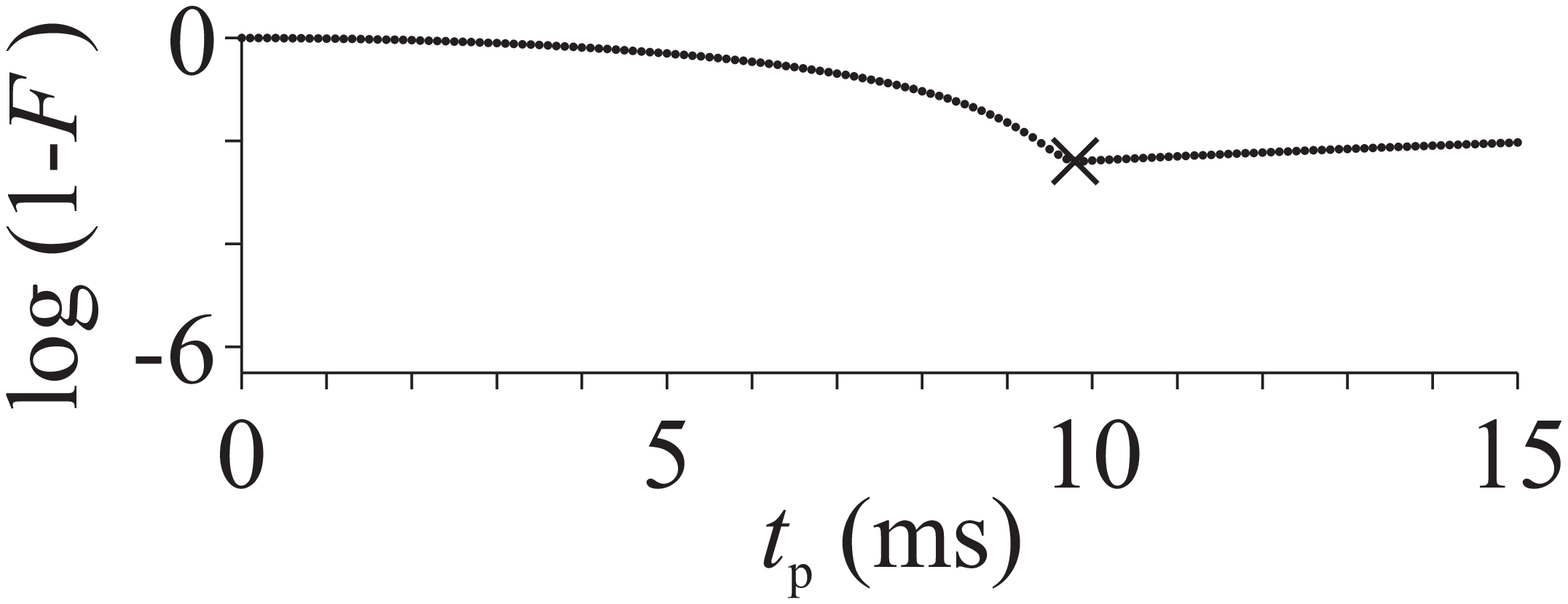}\\
		\raisebox{0.6cm}{6} &
		\includegraphics[width=0.45\columnwidth]{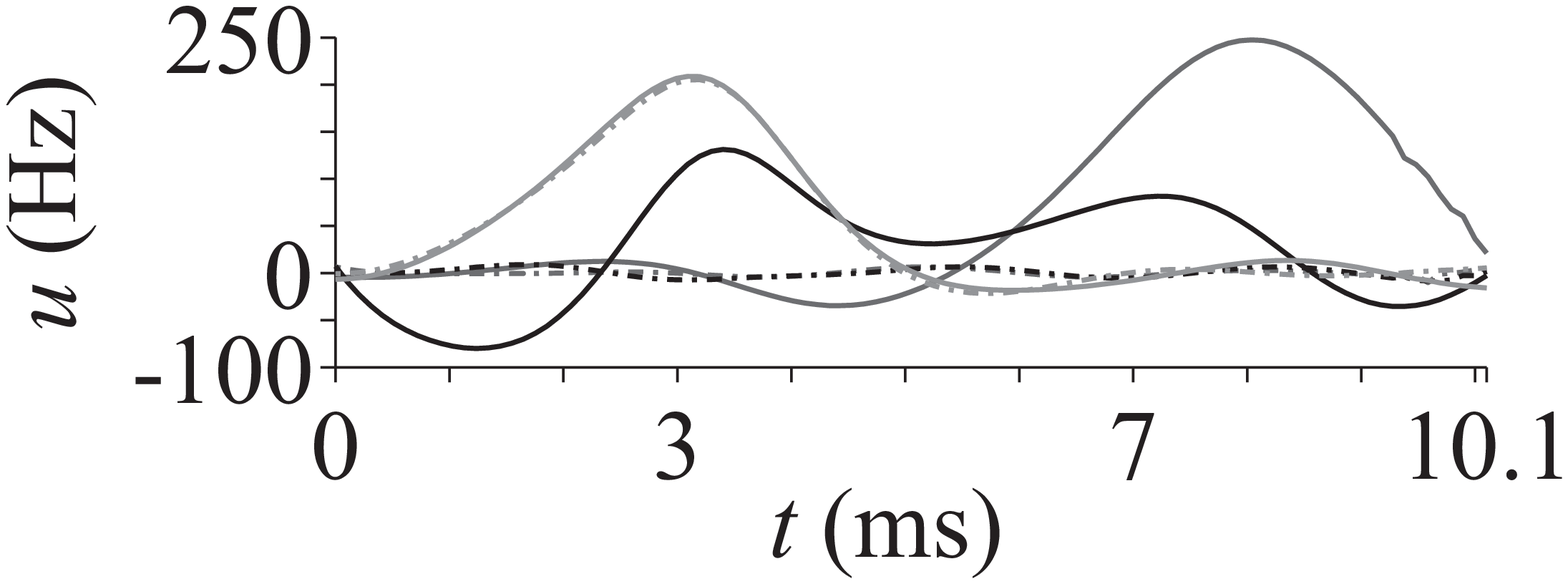} &
		\includegraphics[width=0.45\columnwidth]{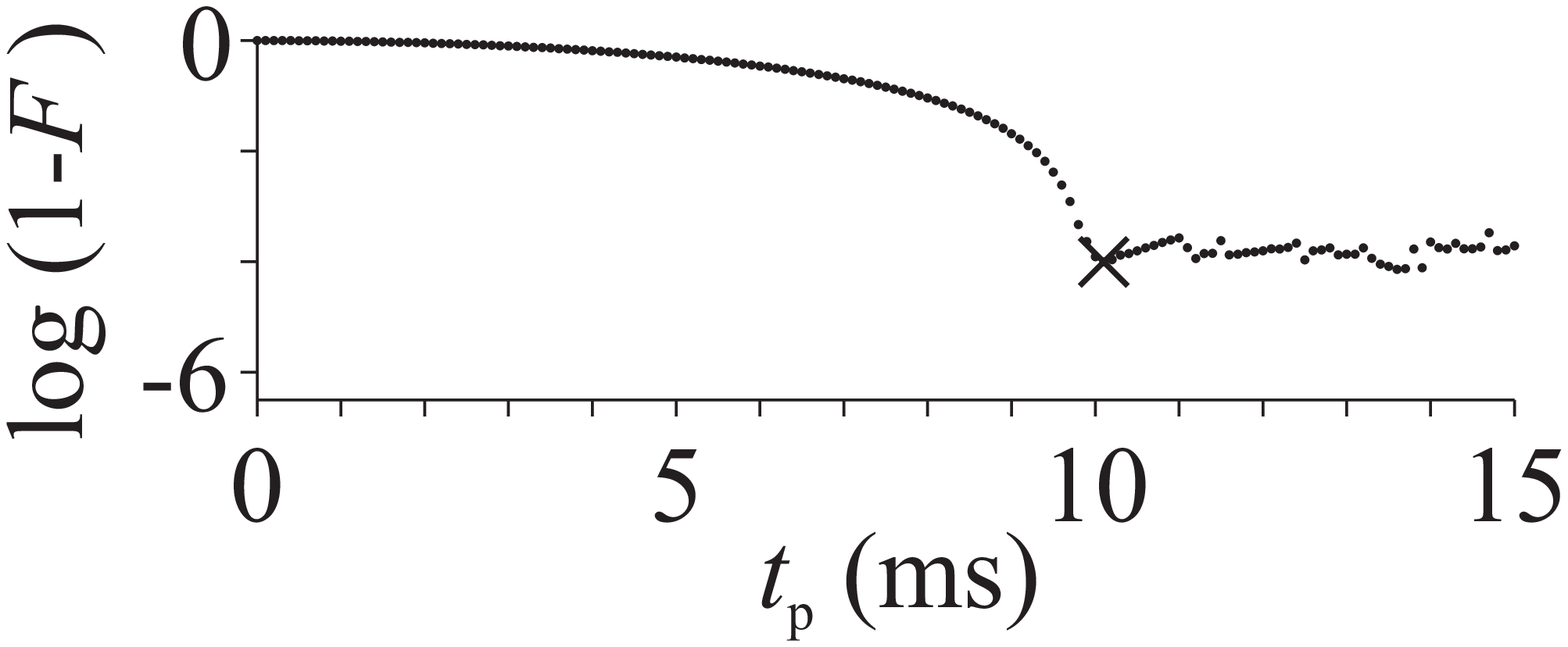}\\\\[-3mm]
		\hline \\[-3.5mm]
		\multicolumn{3}{l}{(b)~$k=1.59$:}\\
		 \raisebox{0.6cm}{1} &
		\includegraphics[width=0.45\columnwidth]{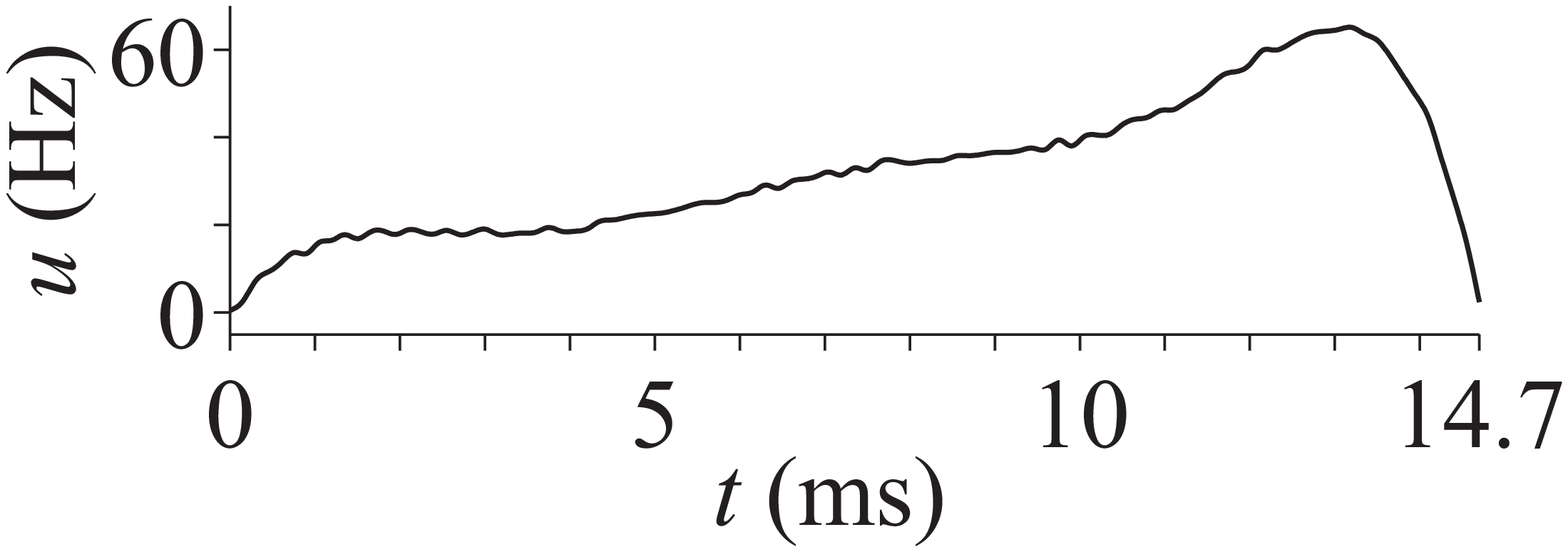} &
		\includegraphics[width=0.45\columnwidth]{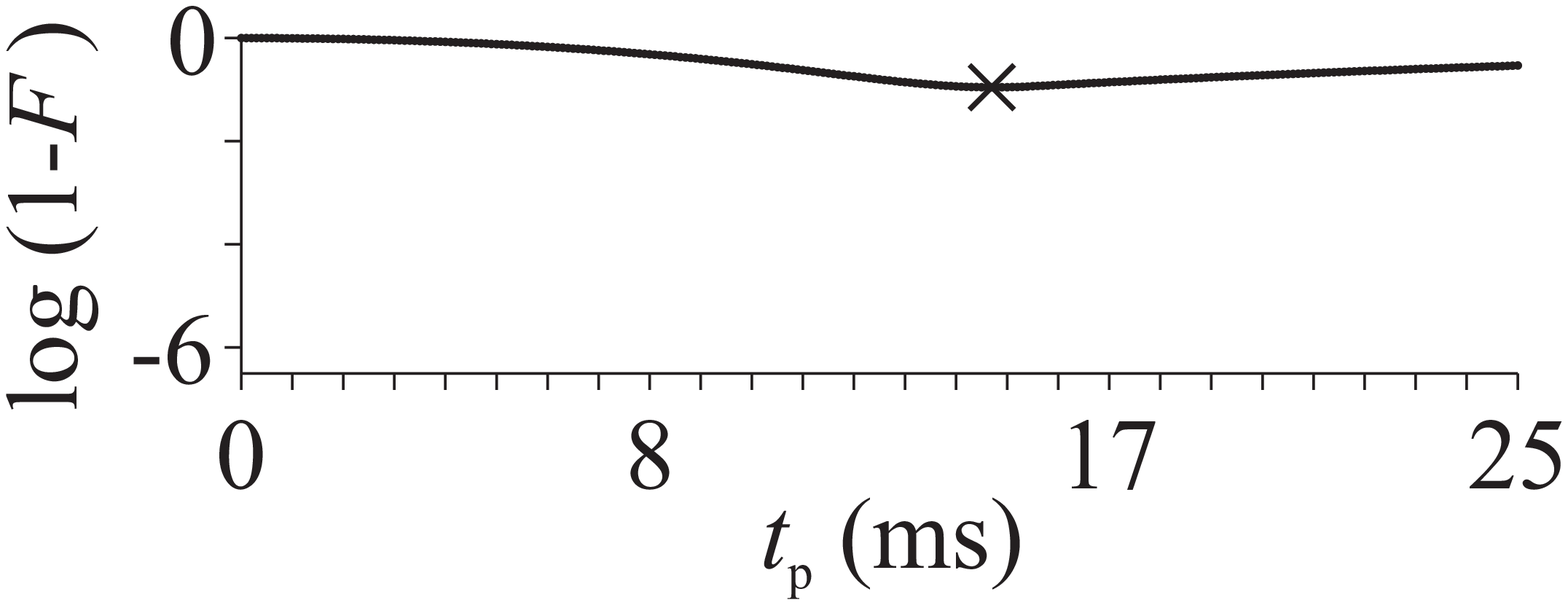}\\
		\raisebox{0.6cm}{6} &
		\includegraphics[width=0.45\columnwidth]{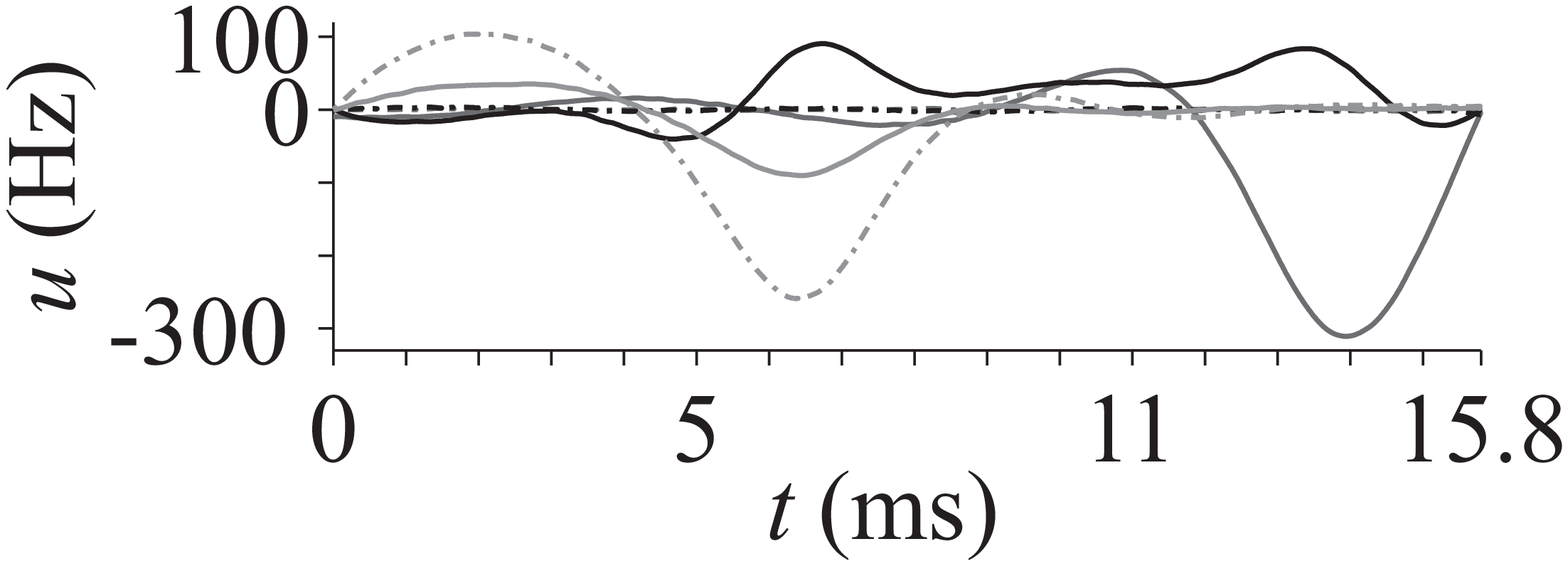} &
		\includegraphics[width=0.45\columnwidth]{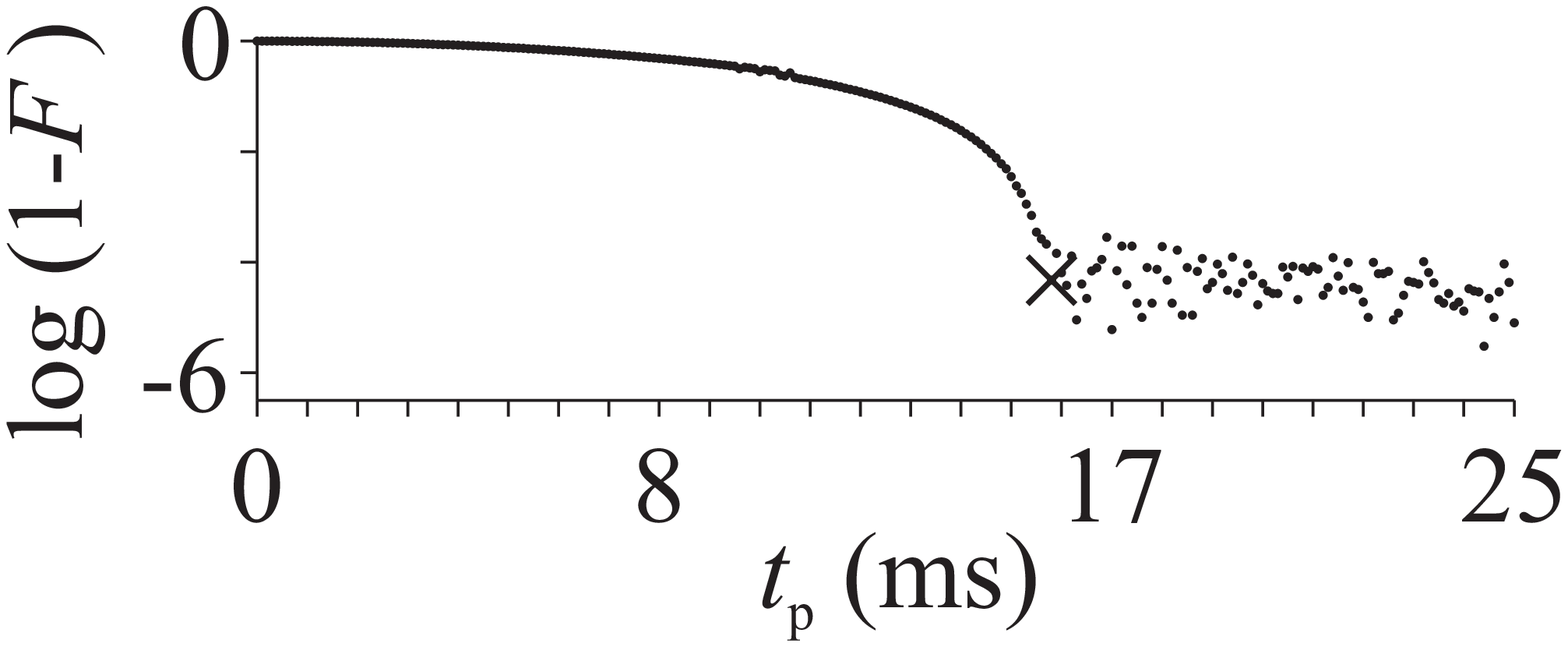}\\
		\end{tabular}
	\end{ruledtabular}
\end{table}

\subsection{Four-spin system\label{sec:3_4_spin_loopedspin_4spin}}

Following Section~\ref{sec:3_4_spin_numeric}, we numerically optimize
the shaped pulses for a more generally-coupled four-spin systems. 
Analyzing the numerical results (see Table~\ref{tab:loop_4spin_k241_shape}), 
we can say that this system needs all eight rf controls on each spin along 
both the $x$- and $y$-direction in order to achieve the coherence transfer 
with minimum duration and maximal fidelity. Table~\ref{tab:tab2} summarizes 
and compares the duration $t_p$ and fidelity $F$ of shaped pulses for more 
generally-coupled spin systems.

\begin{table}[tb]
\caption{\label{tab:loop_4spin_k241_shape}%
Numerical results for a more generally-coupled four-spin systems with coupling ratios $k_1=2.38 \approx J_{12}/J_{23}$ and
$k_2=0.94 \approx J_{34}/J_{23}$ (i.e.\ $J_{12}=46~\text{Hz}$, $J_{23}=19.3~\text{Hz}$, and $J_{34}=18.1~\text{Hz}$)  as well as additional couplings $J_{13}=4.1~\text{Hz}$ and $J_{24}=2~\text{Hz}$. Compare to
Table~\ref{tab:linear_4spin_k1_shape}.}
	\begin{ruledtabular}
		\begin{tabular}[t]{c  c    c }
		$\#{}u$ & pulse shape & logarithmic TOP curve\\
		\hline \\[-3mm]
		 \raisebox{0.6cm}{2} &
		\includegraphics[width=0.45\columnwidth]{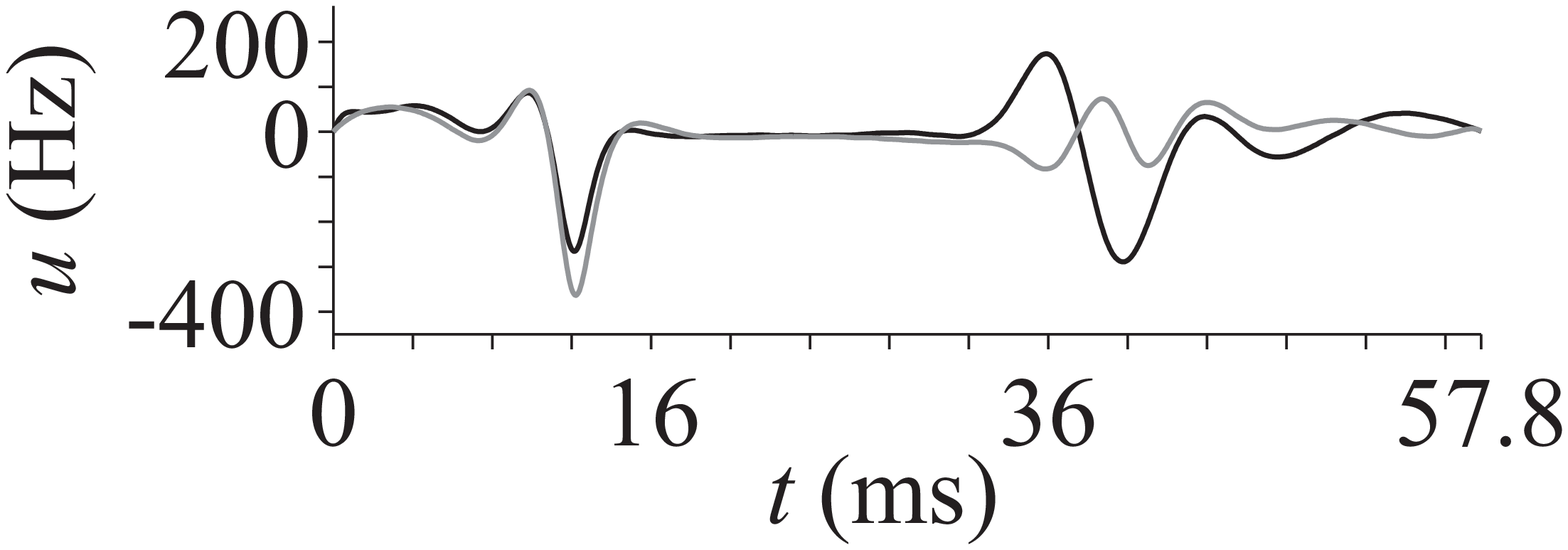} &
		\includegraphics[width=0.45\columnwidth]{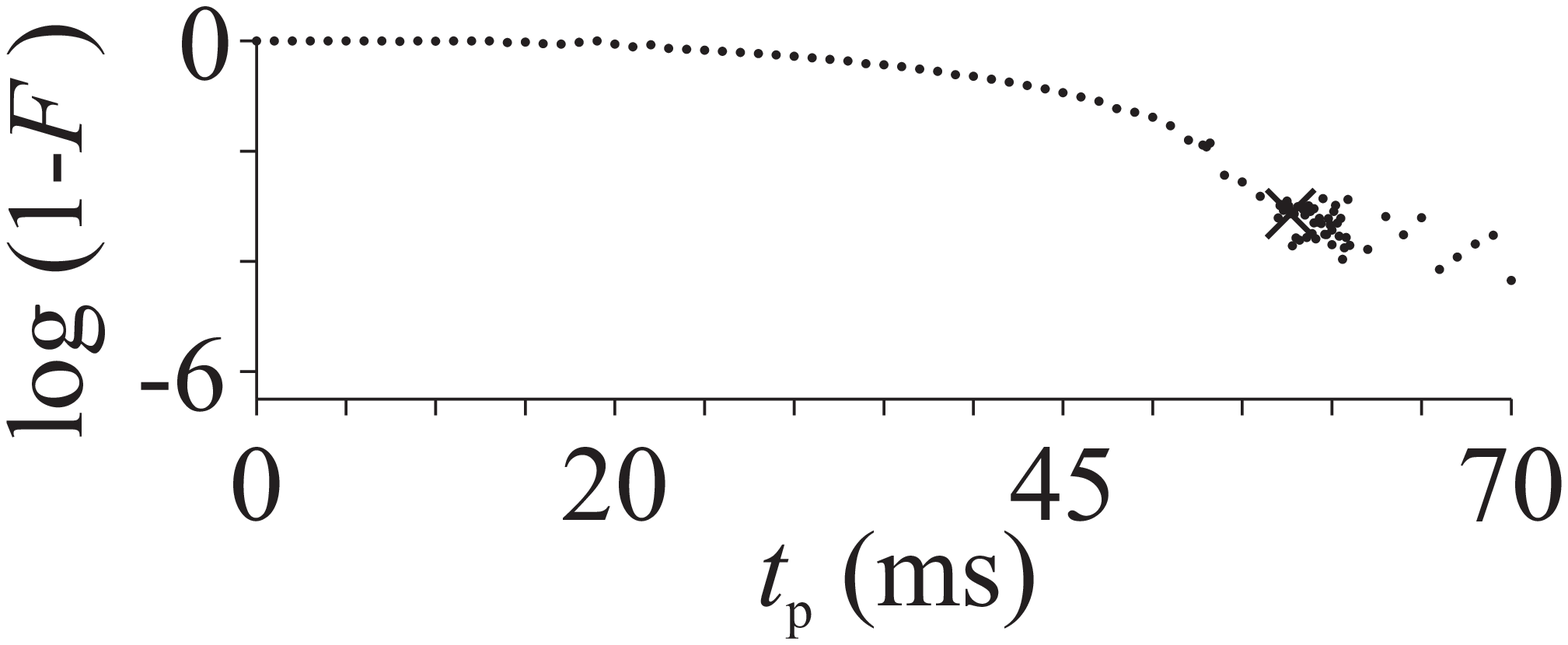}\\
		\raisebox{0.6cm}{8} &
		\includegraphics[width=0.45\columnwidth]{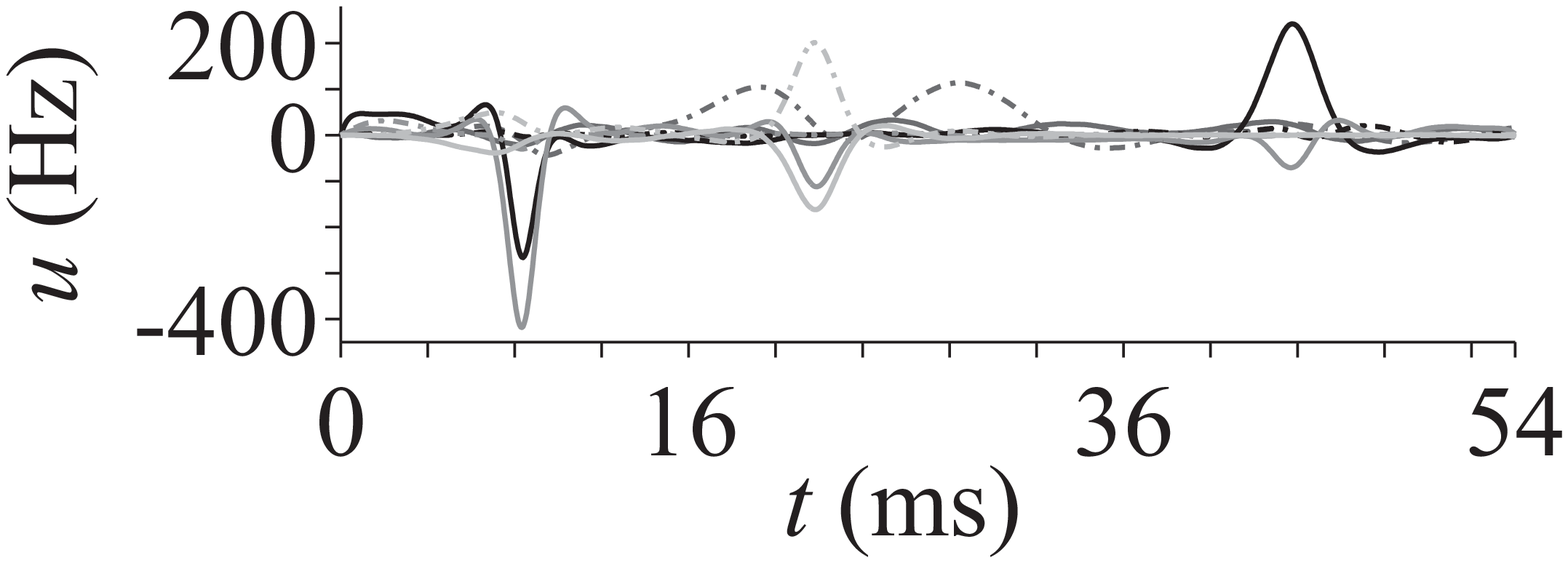} &
		\includegraphics[width=0.45\columnwidth]{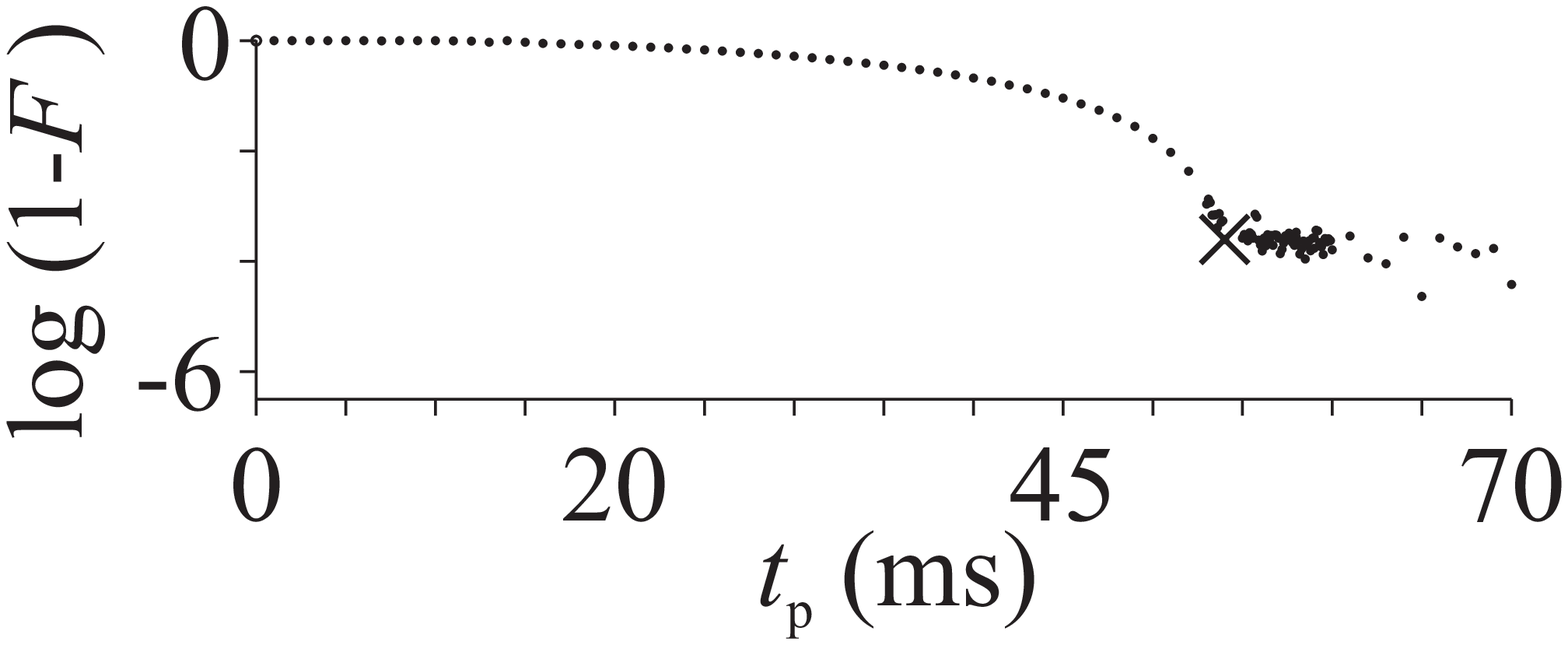}\\
		\end{tabular}
	\end{ruledtabular}
\end{table}

\begin{table}[tb]
\caption{\label{tab:tab2}%
We compare the duration $t_p$ and fidelity $F$ of numerically-optimized
shaped pulses in the case of three- and four-spin systems allowing a varying number 
of rf controls $u$. Using only one rf control
we show the effect of indirect couplings---which are usually
present in experiments---on the  
fidelity of optimized pulses.  
Hence more rf controls are necessary for 
higher fidelities. The $J$-values are taken 
from the actual spin systems shown 
in Figs.~\ref{fig:3_quibit} and \ref{fig:5_quibit}.
}
	\begin{ruledtabular}
		\begin{tabular}{c c c c c c}
			\textrm{graph}&
			\textrm{$J_{13}$~(Hz)}&
			\textrm{$J_{24}$~(Hz)}& 
			\textrm{$\#u$}&
			\textrm{$t_p$~(ms)}&
			\textrm{$F$}\\
			\hline
\multirow{8}{*}{\includegraphics[width=0.24\columnwidth]{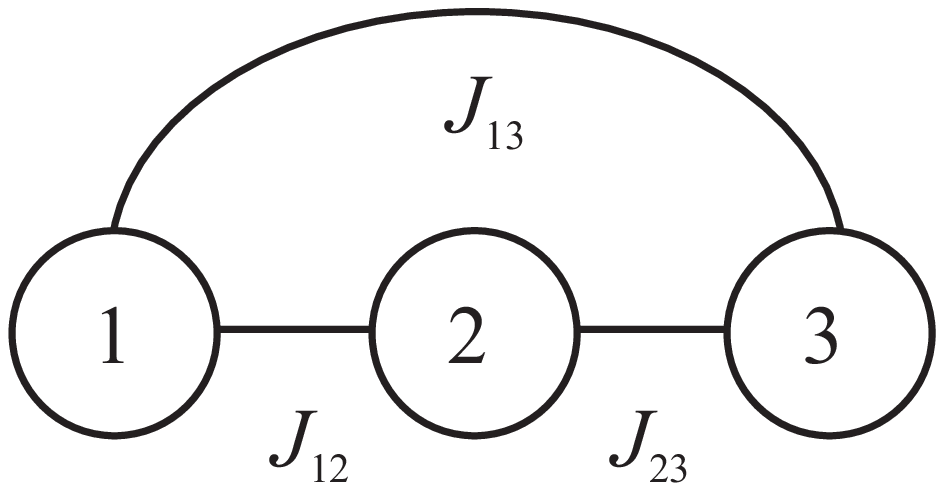}}
			& \multicolumn{5}{l}{$k=1$}\\
			& 0.0 & -- & 1 & 9.8 & 0.9999\\
			& 2.9 & -- & 1 & 9.8 & 0.9959\\
			& 2.9 & -- & 6 & 11.5 & 0.9999\\
			\cline{2-6}
			& \multicolumn{5}{l}{$k=1.59$}\\
			& 0.0 & -- & 1 & 15.5 & 0.9999\\ 
			& 10.0 & -- & 1 & 15.5 & 0.8837\\
			& 10.0 & -- & 6 & 15.8 & 0.9999\\
		\hline
\multirow{5}{*}{\includegraphics[width=0.34\columnwidth]{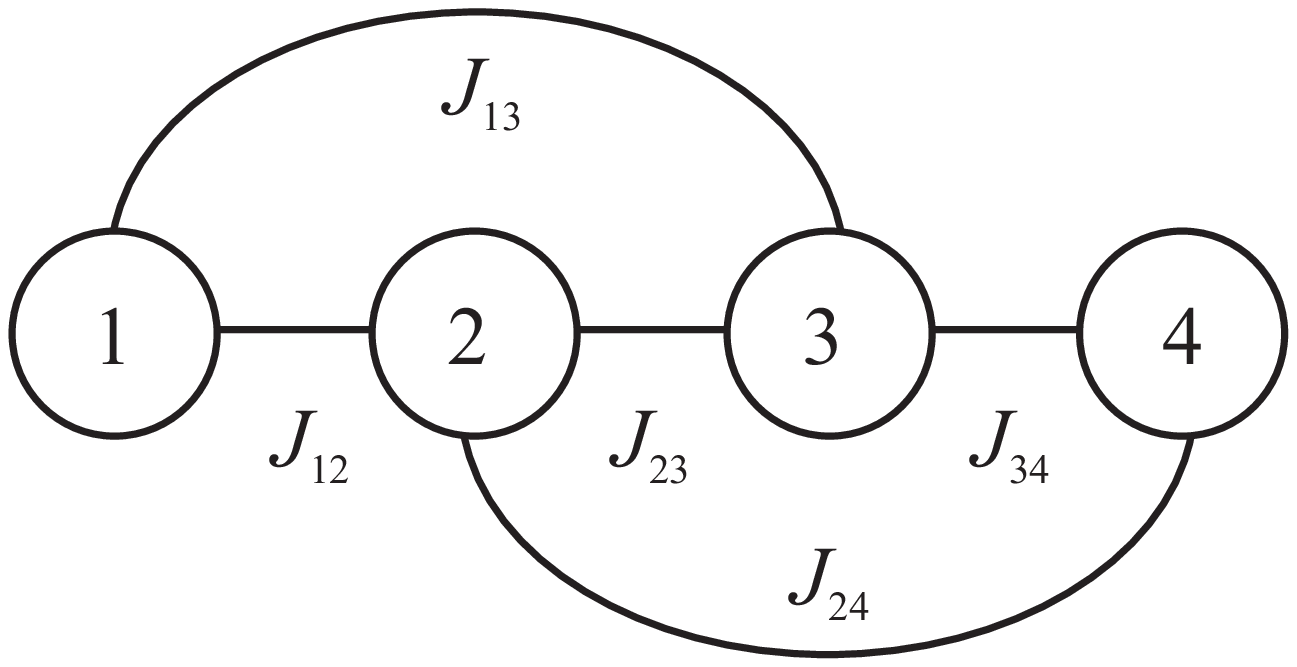}}
			 & \multicolumn{5}{l}{$k_1=2.38$~and~$k_2=0.94$}\\ 
			 & 0.0 & 0.0 & 2 & 53.2 & 0.9999\\ 
			 & 4.1 & 2.0 & 2 & 53.2 & 0.9859\\
			 & 4.1 & 2.0 & 8 & 54.0 & 0.9999\\[0mm]
		\end{tabular}
	\end{ruledtabular}
\end{table}

\section{Experimental results\label{sec:experiment}}

\begin{figure}[tb]
\includegraphics[width=0.75\columnwidth]{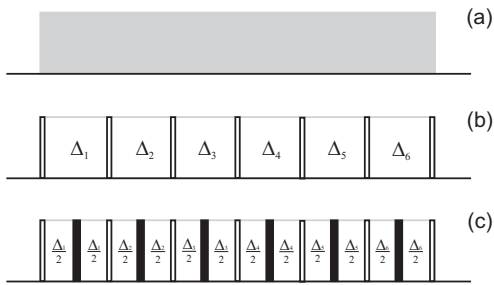}
\caption{In the DANTE approach an on-resonance shaped pulse (see~(a)) 
is converted to a series of short hard pulses and delays $\Delta_{i}$ 
(see~(b)). Then, the pulse can be converted to a broadband pulse by 
inserting a refocusing element (i.e.\ $\pi$-pulse) represented by solid bars
between two hard pulses (see~(c)).}
\label{fig:dante_approach}
\end{figure}

\begin{figure}[tb]
\includegraphics[width=0.85\columnwidth]{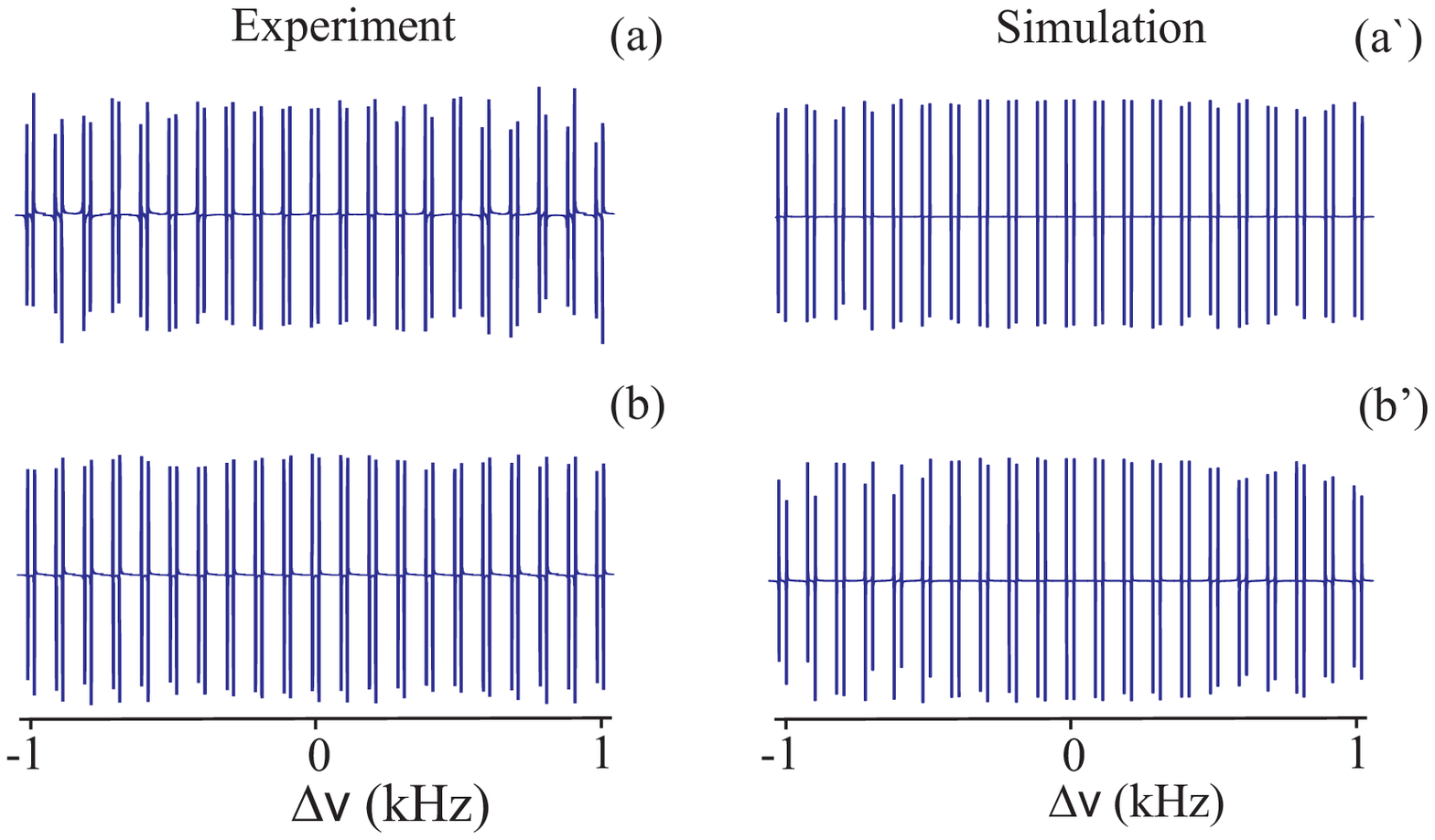}
\caption {(Color online) We compare for a three-spin system the offset ($\Delta \sf{v}$) 
profile for $\pm 1$~kHz of the anti-phase signal (see text)
resulting from a conventional pulse sequence ($t_p=11.4~\text{ms}$) 
in the case of experiment 
(see~(a)) and simulation (see~(a')) with broadband versions of the 
analytic pulses ($t_p=9.8~\text{ms}$, see~(b) and (b')) for the case of coupling ratio $k=1$
of three spins.}
\label{fig:k1_exp_sim_main}
\end{figure}

\begin{figure}[tb]
\includegraphics[width=0.95\columnwidth]{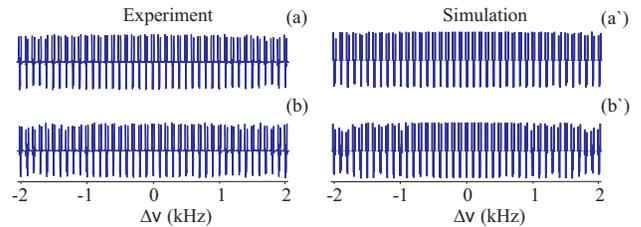}
\caption {(Color online) We compare for a three-spin system the  offset ($\Delta \sf{v}$) 
profile for $\pm 2$~kHz of the anti-phase signal (see text)
resulting from a conventional pulse sequence ($t_p=17.7~\text{ms}$) 
in the case of experiment 
(see~(a)) and simulation (see~(a')) with broadband versions of the 
analytic pulses ($t_p=15.5~\text{ms}$, see~(b) and (b')) for the case of coupling ratio $k=1.59$.}
\label{fig:k159_exp_sim_main}
\end{figure}

Analytic and numerically-optimized pulses are usually optimized for
on-resonance cases. We follow the DANTE approach \cite{KLKLG04,MF78,CBR95} in order to obtain 
pulses which are broadband, i.e. invariant with respect to the change 
of the chemical shift in a given offset range (Fig.~\ref{fig:dante_approach}). 
First, a shaped pulse is converted into a sequence of short hard pulses and delays.
We used hard pulses with constant flip angles (see below), and the delays between 
the hard pulses correspond to the time required by the shaped pulse to accumulate this flip angle \cite{KLKLG04,MF78,CBR95}.
Then, a refocusing element (i.e.\ $\pi$-pulse) 
\cite{KLKLG04,KHSY:2007} is inserted between two hard pulses. 
The offset
bandwidth covered by a refocused DANTE sequence is directly proportional to the rf 
amplitude of the hard pulses and the $\pi$-pulses used in the sequence.

All the experiments are implemented on a Bruker AVANCE III 600~MHz spectrometer 
at 298 Kelvin: We use a triple resonance TXI probe head with Z-gradient in the 
case of the three-spin system with $k = 1$. 
For the three spin system with $k = 1.59$ and
the four spin system with $k_1 = 2.38$ and $k_2 = 0.94$ we use
a custom-made 6-channel probe head with Z-gradient  
addressing all nuclei ${{}^{19}}\mathrm{F}$, ${{}^{1}}\mathrm{H}$,
${{}^{31}}\mathrm{P}$, ${{}^{12}}\mathrm{C}$ (or ${{}^{13}}\mathrm{C}$), and ${{}^{14}}\mathrm{N}$ (or ${{}^{15}}\mathrm{N}$)
(see \cite{pomplun:2010,MPBZEFG11}).
In the experiments for three spins we use the molecules shown in 
Fig.~\ref{fig:3_quibit}. The experiment for the coupling ratio $k=1$ 
uses the first molecule (see Fig.~\ref{fig:3_quibit}(a)) which is 
dissolved in deuterated water $\mathrm{D}_2\mathrm{O}$. For $k=1.59$ 
we use the second molecule (see Fig.~\ref{fig:3_quibit}(b)) dissolved 
in deuterated methanol $\mathrm{C}\mathrm{D}_3\mathrm{O}\mathrm{D}$. 
The simulated and experimental 
offset profiles are shown in 
Figs.~\ref {fig:k1_exp_sim_main} and \ref{fig:k159_exp_sim_main}.
We emphasize that the duration of the broadband versions of the analytic or the 
numerically-optimized pulses is shorter than for the conventional pulse 
sequence while keeping its robustness.

We first discuss two three-spin systems:
In the case of $k=1$, we start from the initial polarization $I_{1z}$
of ${^{1}}\mathrm{H}$
(which models the first spin)
and apply a $\tfrac{\pi}{2}$-pulse along $+y$-direction in order
to obtain the coherence $I_{1x}$.
By applying a broadband version of our shaped pulse 
of ${^{15}}\mathrm{N}$ (which models the second spin) 
we get the three-spin coherence $4I_{1y}I_{2y}I_{3z}$. 
The broadband version of this shaped pulse is divided into four hard pulses with an amplitude of $4145.936~\text{Hz}$, a flip angle of $45.00$~degrees,
and zero phase; 
it also contains refocusing $\pi$-pulses where the phases
are chosen according the MLEV-4  cycle \cite{LFF82}. 
Next, we apply $\tfrac{\pi}{2}$-pulse on  ${^{15}}\mathrm{N}$ along the 
$x$-direction and we obtain the coherence $4I_{1y}I_{2z}I_{3z}$. In the end, we can 
detect an anti-phase signal of ${^{1}}\mathrm{H}$ (first spin) with respect to the spins of ${^{15}}\mathrm{N}$ and ${^{1}}\mathrm{H}$ (which models the third spin).
Similarly, 
in the case of $k=1.59$, we start from the initial coherence $I_{1z}$
on the spin of ${^{1}}\mathrm{H}$ (which models the first spin)
and apply a $\tfrac{\pi}{2}$-pulse along $+y$-direction in order
to obtain the coherence $I_{1x}$.
Then, we apply the broadband version of our shaped 
pulse on the spin of  ${^{19}}\mathrm{F}$ in order to produce the three-spin coherence
$4I_{1y}I_{2y}I_{3z}$. 
The broadband version of this shaped pulse is divided into four hard pulses with an amplitude of $10000~\text{Hz}$,
a flip angle of $45.03$~degrees,
and zero phase; it also contains refocusing $\pi$-pulses where the phases
are chosen according the MLEV-4  cycle \cite{LFF82}. 
In the next step, we apply a 
$\tfrac{\pi}{2}$-pulse on the spin of ${^{1}}\mathrm{H}$ and we end 
up with the coherence $4I_{1z}I_{2y}I_{3z}$. Finally, we detect an anti-phase signal 
on the spin
of ${^{19}}\mathrm{F}$ with respect to the spins of ${^{1}}\mathrm{H}$ 
and ${^{31}}\mathrm{P}$.

\begin{figure}[tb]
\includegraphics[width=0.95\columnwidth]{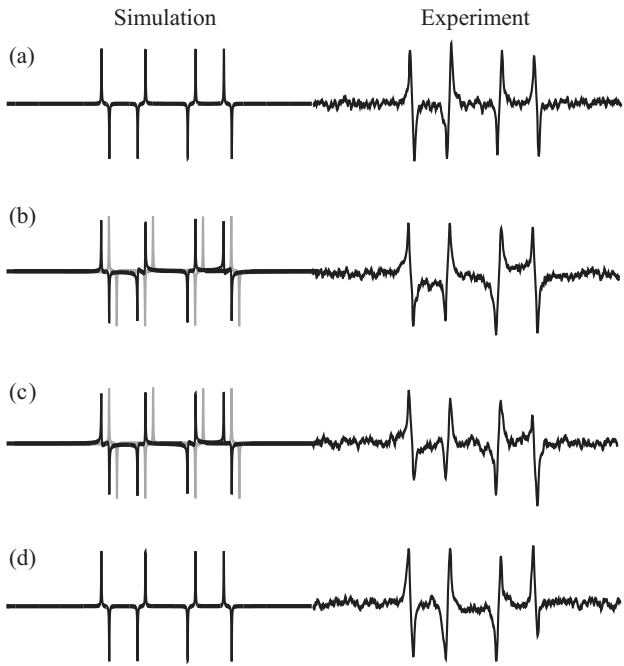}
\caption {We show the 
anti-phase signal of the spin of ${^{19}}\mathrm{F}$ with respect to the spins of 
${^{1}}\mathrm{H}$, ${^{13}}\mathrm{C}$, and ${^{15}}\mathrm{N}$ 
of a four-spin system
corresponding to
the simulation (left) and experiment (right). 
We use 
the conventional pulse sequence (a) ($t_p=64.4~\text{ms}$), an analytical pulse 
sequence (b) ($t_p=53.9~\text{ms}$), a pulse which was numerically-optimized
for the abstract linear spin chain with rf controls on 
the second and third spin (c) ($t_p=53.2~\text{ms}$), and a pulse which was numerically-optimized 
for the more generally-coupled spin system with rf controls on all spins (d) ($t_p=54.0~\text{ms}$).
The simulation for the abstract linear spin chain is given in gray color. 
All the other plots for the more realistic case of a more generally-coupled 
spin system are given in black color.
The plots are scaled vertically  by a factor of two.}
\label{fig:4spin_sim_exp}
\end{figure}

In the four-spin system, we show on-resonance simulations and 
experiments for numerically-optimized shaped pulses
comparing the conventional approach with analytical 
and numerically-optimized pulses (see Fig.~\ref{fig:4spin_sim_exp}).
The corresponding experiments are implemented on the
molecule of Fig.~\ref{fig:5_quibit},
which is dissolved in deuterated acetonitrile.
Figures~\ref{fig:4spin_sim_exp}(b) and (c) show a reduction in signal 
intensity for the simulation if we compare the effect of the pulse on the abstract linear spin chain (shown in gray)
with the effect on  the more realistic and more generally-coupled spin system (shown in black) as
the corresponding pulses were only optimized for the abstract linear spin chain.
We remark that the pulse
of Fig.~\ref{fig:4spin_sim_exp}(d) is optimized for a more generally-coupled spin system
while using rf-controls on all spins.
Thus, we conclude---using also the data of Table~\ref{tab:tab2}---that
the pulse
of Fig.~\ref{fig:4spin_sim_exp}(d) 
shows a higher fidelity when compared to the pulses of 
Figs.~\ref{fig:4spin_sim_exp}(b) and (c).
Furthermore, the pulse corresponding to Fig.~\ref{fig:4spin_sim_exp}(d)
is shorter (by $14\%$) than the conventional pulse sequence corresponding to Fig.~\ref{fig:4spin_sim_exp}(a)
while maintaining its robustness to additional couplings (see also Table~\ref{tab:tab3}).

\section{Linear spin chains with more than four spins\label{sec:longer_linspin}}

\begin{figure}[tbh]
\includegraphics[width=0.49\columnwidth]{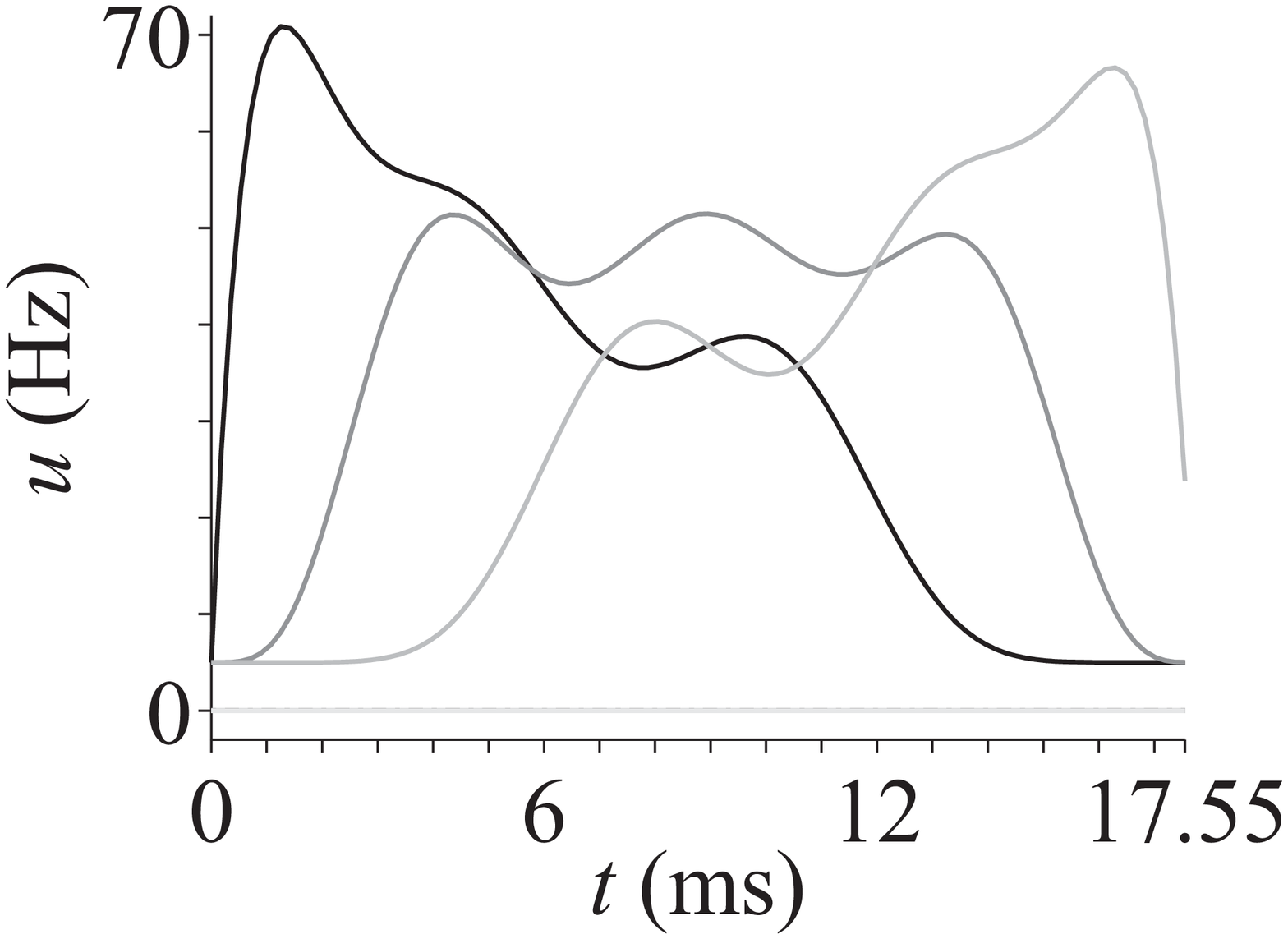}  
\includegraphics[width=0.49\columnwidth]{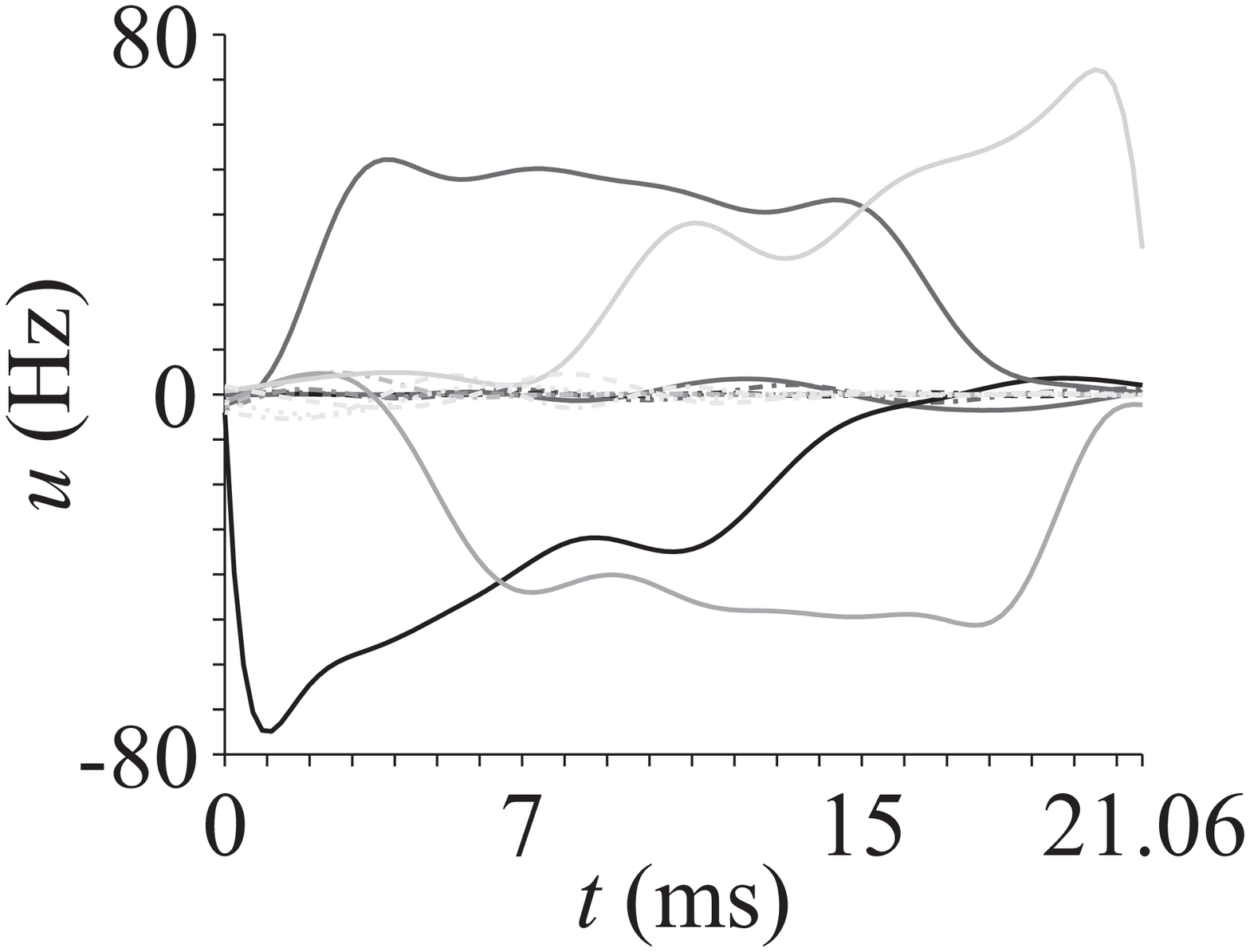}
\caption{Using all ten (or twelve) rf controls we determined numerically-optimized 
pulse shapes for linear spin chains of length five (or six) in the case of $k=1$. 
We remark that most control strengths are very small.}
\label{fig:long_linear_spin_shape}
\end{figure}

In this section, we generalize the numerical optimization of shaped 
pulses to linear spin chains of five and more spins. 
Figure~\ref{fig:long_linear_spin_shape} shows two examples of the 
optimized pulse shapes with coupling ratios $k_\ell=1$ and coupling 
strengths $J_{\ell,\ell+1}=88.05~\text{Hz}$. These examples suggest that 
time-optimal controls can be obtained on multiple spins even  
while irradiating only on the spins two to $\ell-1$ along 
the $y$-direction (cp.\ Sec.~IV of Ref.~\cite{YZK08}). We obtain shorter 
pulses for the numerically-optimized 
pulses compared to the conventional pulse sequences as 
summarized in Table~\ref{tab:tab3}.

\begin{table}[tbh]
\caption{\label{tab:tab3}%
We compare the minimum time $t_p$ required for a coherence transfer 
by numerically-optimized (oc) and conventional (conv) pulse sequences 
for different number $n$ of spins and coupling ratios $k$.}
	\begin{ruledtabular}
		\begin{tabular}{c c c c c c c}
		 \multirow{3}{*}{$n$} & \multicolumn{6}{ c }{$t_p~(s)$} \\ 
		\cline{2-7}
		& \multicolumn{3}{ c }{$k=1$} & \multicolumn{3}{ c }{$k\not= 1\footnotemark[1]$} \\
		\cline{2-7}
		& oc & conv & oc$/$conv & oc & conv & oc$/$conv \\
		\hline
		3 & 0.0098 & 0.0114 & 0.8596 & 0.0155 & 0.0177 & 0.8757 \\
		4 & 0.0138 & 0.0170 & 0.8118 & 0.0532 &	0.0644 & 0.8261 \\
		5 & 0.0177 & 0.0227 & 0.7797 & -- & -- & -- \\
		6 & 0.0216 & 0.0284 & 0.7605 & -- & -- & -- \\
		\end{tabular}
	\end{ruledtabular}
\footnotetext[1]{For $n=3$ we have $k=1.59$. And for $n=4$ we have $k_1=2.38$ and $k_2=0.94$.}
\end{table}

\section{Conclusion\label{sec:conclusion}}
In the case of linear three-spin chains we reproduced numerically the previous analytical results
\cite{YGK07,YZK08} obtaining the same family of restricted controls by applying
pulses only on the second spin along the $y$-axis. The same holds for linear four-spin chains where
we also obtain the analytical family of restricted controls by applying pulses only
on the second and third spin along the $y$-axis; but the numerically-optimized pulses appear to
be a little shorter than the analytical ones. Both for three and
four spins no gain in pulse duration is found if arbitrary pulse structures are allowed.
These observations are summarized in Conjectures~\ref{conj1} and \ref{conj2}.
Even for longer spin chains (consisting of up to six coupled spin-$1/2$) there is some
numerical evidence suggesting that the same restricted controls
motivated by Refs.~\cite{YGK07,YZK08} lead to time-optimal pulses (in the case of unrestricted controls)
for linear spin chains of arbitrary length.

Further numerical results are presented for more general and more realistic coupling topologies, for which
so far no analytical results are known. Compared to linear spin chains we obtain different pulse structures depending on the number of available controls. 
We hope that the presented results and conjectures will motivate further analytical work
in order to develop a better understanding of time-optimal control sequences for the generation
of multi-spin coherence.

Note that the minimum times for the transfers
$
 I_{1\delta}      \rightarrow 
 2^{n-1} I_{1\epsilon_1}   \cdots  I_{n \epsilon_n}  
$
and
$
 2^{n-1} I_{1\epsilon_1}   \cdots  I_{n \epsilon_n}     \rightarrow  
 I_{1\delta} 
$
are identical
($\delta, \epsilon_k  \in \{x,  y,  z  \}$ for $k=1, \ldots ,n$), which is directly relevant for ``out and back" experiments and the reconversion of multiple-quantum coherence to detectable single quantum operators.
In the experimental part, we demonstrated that the optimized pulse sequences work in realistic settings
under relaxation and experimental imperfections (e.g.\ 
inhomogeneity of the control field, miscalibrations, and phase transients).
In addition, the pulses can be made broadband (i.e. robust with respect to frequency offsets) using the DANTE approach.

Here we assumed for simplicity that each spin-$1/2$ can be selectively addressed, which is directly relevant 
to heteronuclear
spin systems but the optimal transfer scheme can also be adapted to homonuclear spin systems.
The presented sequences can be directly applied to small molecules and peptides,
which is in particular true for the broadband versions. The minimum pulse sequence durations  for complete transfer are reduced
by up to $24\%$ compared to conventional approaches (see Table~\ref{tab:tab3}).
Conversely, for a fixed transfer time significantly improved transfer amplitudes are possible, e.g., 
for a linear three-spin chain we gain approximately $23\%$ in transfer efficiency 
when we allow only for half of the transfer time necessary for a complete transfer
(cp.\ Fig.~7).
For large proteins, further gains in efficiency are expected if relaxation-optimized
pulse sequences can be developed for the specific relaxation 
super operator given in the system. Although such sequences are beyond the scope
of the present paper, the results on time-optimal sequences presented here provide
an important benchmark for relaxation-optimized sequences.

\begin{acknowledgments}
MN would like to thank the TUM Graduate school.
RZ is supported by the Deutsche Forschungsgemeinschaft through the grant {\sc schu}~1374/2-1.
SJG acknowledges support from the DFG ({\sc gl}~203/6-1), SFB 631, the EU program Q-ESSENCE, 
and the Fonds der Chemischen Industrie.
We acknowledge the support by the Bayerisches NMR Zentrum,  M\"unchen.
\end{acknowledgments}

%

\end{document}